\documentclass[twocolumn]{aastex63}

\usepackage{amsmath, amsthm, amssymb,amsbsy}
\usepackage{pbox}
\usepackage{booktabs,tabularx}
\usepackage{multirow}
\usepackage{graphicx}
\usepackage{makecell}
\usepackage[flushleft]{threeparttable}
\usepackage{enumitem}

\usepackage[utf8]{inputenc}
\usepackage[T1]{fontenc}
\usepackage{array}
\newcolumntype{P}[1]{>{\centering\arraybackslash}p{#1}}
\newcolumntype{M}[1]{>{\centering\arraybackslash}m{#1}}

\accepted{}
\submitjournal{ApJ}

\shorttitle{Proto-atmosphere Accretion on Eccentric Planets}
\shortauthors{Mai et al.}

\begin{document}

\title{The Dynamic Proto-atmospheres around Low-Mass Planets with Eccentric Orbits}

\correspondingauthor{Chuhong Mai}
\email{chuhong.mai@asu.edu}

\author[0000-0002-9243-5065]{Chuhong Mai}
\affiliation{School of Earth and Space Exploration, Arizona State University, Tempe, AZ 85287, USA}

\author[0000-0002-1571-0836]{Steven J. Desch}
\affiliation{School of Earth and Space Exploration, Arizona State University, Tempe, AZ 85287, USA}

\author[0000-0003-2309-8963]{Rolf Kuiper}
\affiliation{Institut f\"{u}r Astronomie und Astrophysik, Universit\"{a}t T\"{u}bingen, Auf der Morgenstelle 10, 72076 T\"{u}bingen, Germany}

\author[0000-0002-2919-7500]{Gabriel-Dominique Marleau}
\affiliation{Institut f\"{u}r Astronomie und Astrophysik, Universit\"{a}t T\"{u}bingen, Auf der Morgenstelle 10, 72076 T\"{u}bingen, Germany}
\affiliation{Physikalisches Institut, Universit\"{a}t Bern, Gesellschaftsstr. 6, 3012 Bern, Switzerland}
\affiliation{Max-Planck-Institut f\"{u}r Astronomie, K\"{o}nigstuhl 17, 69117 Heidelberg, Germany}

\author[0000-0002-7078-5910]{Cornelis Dullemond}
\affiliation{Institut f\"{u}r Theoretische Astrophysik (ITA), Zentrum f\"{u}r Astronomie (ZAH), Universit\"{a}t Heidelberg, Albert-Ueberle-Str.2 69120 Heidelberg, Germany}

\begin{abstract}

Protoplanets are able to accrete primordial atmospheres when embedded in the gaseous protoplanetary disk. The formation and structure of the proto-atmosphere are subject to the planet--disk environment and orbital effects.  Especially, when planets are on eccentric orbits, their relative velocities to the gas can exceed the sound speed. The planets generate atmosphere-stripping bow shocks. We investigate the proto-atmospheres on low-mass planets with eccentric orbits with radiation-hydrodynamics simulations.  A 2D radiative model of the proto-atmosphere is established with tabulated opacities for the gas and dust. The solutions reveal large-scale gas recycling inside a bow shock structure. The atmospheres on eccentric planets are typically three to four orders of magnitude less massive than those of planets with circular orbits. Overall, however, a supersonic environment is favorable for planets to keep an early stable atmosphere, rather than harmful, due to the steady gas supply through the recycling flow. We also quantitatively explore how such atmospheres are affected by the planet’s relative velocity to the gas, the planet mass, and the background gas density.  Our time-dependent simulations track the orbital evolution of the proto-atmosphere with the planet--disk parameters changing throughout the orbit. Atmospheric properties show oscillatory patterns as the planet travels on an eccentric orbit, with a lag in phase. To sum up, low-mass eccentric planets can retain small proto-atmospheres despite the stripping effects of bow shocks. The atmospheres are always connected to and interacting with the disk gas. These findings provide important insights into the impacts of migration and scattering on planetary proto-atmospheres.

\end{abstract}


\section{Introduction} \label{sec:intro}
Protoplanets are generally believed to form before the gas in their parent nebulas dissipates and are therefore likely to capture proto-atmospheres from the nebula gas (e.g. \citealt{Goldreich2004, Lee2014}). 
Such proto-atmospheres should resemble the solar nebula in composition -- rich in hydrogen and helium.

Early pioneering work by \cite{Mizuno1978, Mizuno1980a} explored the formation of Jupiter-like gas giants by accreting nebula gas. The core accretion theory (e.g. \citealt{Safronov1969, Perri1974, Mizuno1978, Stevenson1982, Pollack1996}) focused on how accumulation of nebula gas onto protoplanets can trigger runaway accretion when the accreted solid and gas masses are comparable, and typically after the mass of the core reaches the critical core mass ($\sim$ 10 M$_{\oplus}$, \citealt{Venturini2015}). 
Only in recent years has the discovery and characterization of exoplanets revealed the ubiquity of low-density super-Earths and mini-Neptunes (e.g. \citealt{Borucki2010, Fressin2013, Batalha2013}). 
Analyses show they can retain the primordial H$_2$/He atmospheres \citep{Weiss2014, Rogers2015}, likely 0.1--10\% by mass \citep{Lopez2014}, although there are exceptions \citep{Espinoza2016}. This has raised interests in the community to understand how lower-mass planets ($<$ 5 M$_{\oplus}$) accrete and retain proto-atmospheres. 

From an astrobiological point of view, the bioessential elements C, H, O, and N in the solar nebula mostly resided in volatile phases, e.g. H$_2$, H$_2$O, CO/CO$_2$, or N$_2$ \citep{Lodders2003}. Terrestrial water may be attributable to H$_2$O accreted from the nebula, or H$_2$O produced as H$_2$ in the proto-atmosphere reacted with iron oxides (e.g. \citealt{Ikoma2006, Wu2018, Williams2019}). A proto-atmosphere may also have supplied Earth with noble gases (e.g. \citealt{Mizuno1980b, Wu2018, Sharp2019}). Modeling the formation of a planet’s proto-atmosphere is required to assess its geochemistry and habitability.

A number of studies have performed detailed calculations on the accretion of nebula-originated primordial atmospheres onto low-mass planets (e.g. \citealt{Lammer2014, Lee2014, Lee2015, Stokl2015, Ginzburg2015, Stokl2016, Kuwahara2019}), indicating Earth-size planets can acquire an H$_2$/He proto-atmosphere with surface pressure up to 10$^3$ bars. 
Sophisticated hydrodynamics codes have been adopted to solve the detailed atmosphere structures.
\cite{Ormel2015a, Ormel2015b} conducted hydrodynamic simulations to solve an isothermal proto-atmosphere around Earth-like planets in 2D and 3D, respectively. The subsequent work by \cite{Cimerman2017} included radiation transport and opacity treatment to consider radiative cooling. These studies revealed the rotation and recycling behaviors of atmospheric gas around the planet, which might prevent low-mass planets from runaway gas accretion.

\begin{figure}[ht!]
\centering
\hspace*{-0.6cm}
\includegraphics[scale=0.4]{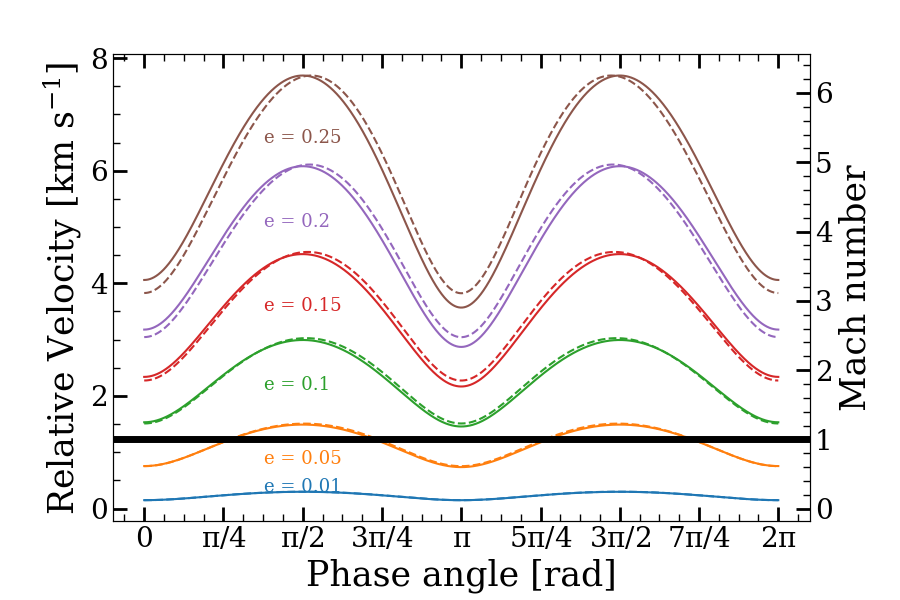}
\caption{The relative velocity between the planet and the disk gas (solid lines), and the corresponding Mach number (dashed lines) throughout orbits with different eccentricities. The Mach number measures a ratio between the flow velocity and the local sound speed. Supersonic flows have Mach numbers larger than unity. The calculations assume a 1 M$_{\oplus}$ planet traveling on an orbit with the semimajor axis as 1 au and no inclination. A higher inclination angle would further increase the relative velocity during midplane passage. Phase angle 0 corresponds to the perihelion of the planet. The thick black line indicates the sound speed in the disk ($\sim$ 300 K) or when Mach number = 1. This value changes very mildly with a realistic radial disk temperature gradient even in the most eccentric case e = 0.25 (e.g. within the thickness of the black line). \label{fig:relative_vel}}
\end{figure}

So far, numerical studies on proto-atmospheres have been focused on planets orbiting their host stars on circular orbits. In particular, ideal spherical symmetry is assumed in 1D simulations by definition to simplify the physics \citep{Stokl2015, Stokl2016}. 
However, protoplanets are not always on circular orbits, especially in young planetary systems. 
Exoplanet discoveries have revealed how common it is for planets to migrate, often stochastically, due to resonances, close encounters or planet--planet scattering (e.g.  \citealt{Chiang2002, Ford2008, Raymond2008, Michtchenko2013, Petrovich2014}). 
To date, many exoplanets have been found on eccentric orbits ($e > 0.1$), regardless of their mass and sizes (e.g. \citealt{Tremaine2004, Antoniadou2015, Kane2016, Wittenmyer2013, Wittenmyer2019}).
In our solar system, Mars is hypothesized to have been scattered from $\sim$ 1 au to an eccentric orbit before settling into its present orbit at $\sim$ 1.5 au \citep{Hansen2009}; and the presence of an ice giant on a very elliptical orbit scattered from $\sim$ 10 au has been hypothesized \citep{Thommes1999, Nesvorny2011} and inferred in observations as the putative “Planet Nine” \citep{Batygin2016}. 
The orbital velocity of a planet or protoplanet on an eccentric orbit will significantly differ from that of the surrounding gas, which remains on circular orbits with approximately the Keplerian speed.
Eccentric planets can strongly disturb the nebular gas and even the background magnetic fields \citep{Mai2018}.
It is unclear how such relative motions between the planet and the gas will change the gas dynamics.
Planets on slightly eccentric orbits may accrete gas faster, while planets on more eccentric orbits (e.g. move supersonically relative to the gas) generate a bow shock which can fundamentally change how it accretes – or loses – a proto-atmosphere. 
In fact, the relative motion between the planet and the gas becomes supersonic throughout the orbit once the eccentricity reaches $\sim$ 0.08 (assuming the planet is 1 au away from the star, see Fig.~\ref{fig:relative_vel}).

Despite the importance of these orbital effects, there have been no numerical models that have considered the implications of eccentric orbits for the presence of proto-atmospheres.
In this work, we investigate how planets on eccentric orbits retain proto-atmospheres, particularly when traveling supersonically through the gas and producing bow shocks.
We aim to address the following questions:
\begin{enumerate}
    \item How does the proto-atmosphere interact with nebula gas when the planet is moving through the disk supersonically? 
    \item What is the flow structure around the planet? How much gas can the planet hold?
    \item How sensitive is the proto-atmosphere to the disk environment and the planet mass?
    \item How does the proto-atmosphere evolve throughout an eccentric orbit? Does the planet's atmosphere gain and lose gas periodically?
\end{enumerate}

The paper is organized as follows: \S~\ref{sec:method} describes the hydrodynamic simulation set-up, including the equations for radiation-hydrodynamics  (\S~\ref{subsec:equations}), the physical and numerical set-up (\S~\ref{subsec:model}), and the list of simulations preformed in this study (\S~\ref{subsec:list}). In \S~\ref{sec:results} we present the simulation results of our canonical set-up (\S~\ref{subsec:canonical}), the variations with different disk and planet properties (\S~\ref{subsec:sensitive}) and the temporal evolution throughout the orbit (\S~\ref{subsec:orbit}). We discuss the findings in \S~\ref{sec:discussions} and summarize the study in \S~\ref{sec:summary}.

\section{Hydrodynamic Simulations} \label{sec:method}
\subsection{Problem set-up and Equations} \label{subsec:equations}
Proto-atmosphere formation on planets with eccentric orbits ($e \geq 0.1$) can be modeled as a gravitating spherical object moving supersonically in the inviscid, compressible and homogeneous gas. 
In this study, we include radiation transport of energy and assume the disk gas as an ideal gas.
We adopt the local thermal equilibrium (LTE) assumption and linearization two-temperature approach\footnote{The two-temperature approach solves for the radiation energy density as well as for the internal gas temperature at the same time, with gas temperature and radiation temperature are treated separately.} for the Flux Limited Diffusion (FLD) approximation for radiation transfer \citep{Kuiper2010, Kuiper2018}.
For convenience, we adopt a frame co-moving with the planet in all simulations.
The planet object is placed in a wind tunnel instantaneously in the beginning and the subsequent gas evolution is simulated using a radiation-hydrodynamics model until a steady/quasi-steady state is reached.

The problem can be studied with sophisticated hydrodynamics codes (e.g. \texttt{PLUTO}, see \S~\ref{subsec:model}) solving the Euler's equations representing the conservation of mass, momentum, and energy:
\begin{equation} \label{Eq1}
    \frac{\partial \rho}{\partial t} + \pmb{\nabla} \cdot (\rho \pmb{v}) = 0 
\end{equation}
\begin{equation} \label{Eq2}
    \frac{\partial}{\partial t} \rho \pmb{v} + \pmb{\nabla} \cdot (\rho \pmb{v}\otimes \pmb{v}) + \pmb{\nabla} P = \rho \pmb{a}_{\text{ext}}
\end{equation}
\begin{equation} \label{Eq3}
    \frac{\partial E}{\partial t} + \pmb{\nabla} \cdot [(E + P) \pmb{v}] = \rho \pmb{v}\cdot \pmb{a}_{\text{ext}}
\end{equation}
where $\rho$ is the gas mass density, $\pmb{v}$ is the relative velocity between the planet and the gas, $P$ is the gas pressure, $\pmb{a}_{\text{ext}}$ is the acceleration caused by external forces, $E = E_{\text{int}} + E_{\text{rad}} + E_{\text{kin}}$ is the total energy density of the gas, which is the sum of internal energy density ($E_{\text{int}} = c_{\text{v}} \rho T$, since we are using a constant $c_{\text{v}}$; $c_{\text{v}}$ is the heat capacity at constant volume, T is gas temperature), radiation energy density ($E_{\text{rad}} = aT^4$, $a$ is the radiation constant), and kinetic energy density ($E_{\text{kin}} = (1/2)\rho v^2$).

With operator splitting we separately solve the transport term $\pmb{\nabla}(E\pmb{v})$, leaving the time derivatives of the internal and radiation energy density as in \citep{Kuiper2010, Cimerman2017}:
\begin{equation} \label{Eq4}
    \frac{\partial (E_{\text{int}} + E_{\text{rad}})}{\partial t} = -\pmb{\nabla} \cdot \pmb{F} - P\pmb{\nabla}\cdot \pmb{v}
\end{equation}
where $\pmb{F}$ is the flux of radiation energy density. Substituting $E_{\text{int}}$ with $c_{\text{v}}\rho T$ and $E_{\text{rad}}$ with $aT^4$ we have:
\begin{equation} \label{Eq5}
    \frac{\partial E_{\text{rad}}}{\partial t} = -\left(\frac{c_{\text{v}}\rho}{4aT^3} + 1\right)^{-1}(\pmb{\nabla} \cdot \pmb{F} + P\pmb{\nabla}\cdot \pmb{v})
\end{equation}
and from FLD approximation:
\begin{equation} \label{Eq6}
    \pmb{F} = - \frac{\lambda c}{\rho \kappa_{\text{R}}}\pmb{\nabla}E_{\text{rad}}
\end{equation}
where $\lambda$ is the flux limiter following the choice in \cite{Levermore1981}, c is the light speed, and $\kappa_{\text{R}}$ is the Rosseland mean opacity. 

Note that the above calculations do not include irradiation from the central planet as the external sources of luminosity from the planet surface (due to radiogenic heat and/or planetesimals accretion) are negligible in our set-up. The heat generated by planetesimals accretion with a mass accretion rate of 10$^{-7}$ M$_{\text{pl}}$/yr \citep{Stokl2015} is three to four orders of magnitude lower than that from gas accretion in our models (M$_{\text{pl}}$ is the planet mass). The radiogenic energy with a reference level of 10$^{21}$ erg s$^{-1}$ M$_{\oplus}^{-1}$ (for early Earth, \citealt{Stacey2008}) is at least two additional orders of magnitude lower \citep{Stokl2015}. For a more detailed analysis of the FLD approximation in radiation transport please refer to \cite{Kuiper2010}.

The external forces considered in the gas system include the gravity from the planet object: 
\begin{equation} \label{Eq7}
    \pmb{a}_{\text{ext}} = -\frac{GM_{\text{pl}}}{r^2}\pmb{e}_{\text{r}}
\end{equation}
where $G$ is the gravitational constant, $M_{\text{pl}}$ is the planet mass and $r$ is the distance to the center of the planet. Following \cite{Cimerman2017}, we do not consider ionization or dissociation of molecules as we focus on solving the flow structure and bulk atmospheric properties. 
In fact, the models in \cite{Desch2002} and \cite{Morris2010} found that $\sim$ 7\% of H$_2$ in nebular shocks can be dissociated, but the total energy of the gas remains unchanged.
The radiative forces are neglected because they are small compared to gravity from the planet and the gas pressure.

The set of Euler's equations (Equ. \ref{Eq1}, \ref{Eq2} and \ref{Eq3}) is closed by the perfect gas equation of state:
\begin{equation} \label{Eq8}
    P = (\gamma - 1)E_{\text{int}} = \frac{\rho}{\mu m_{\text{H}}}k_{\text{B}} T
\end{equation}
where $\gamma$ is the adiabatic index that we fix as $7/5$ (for molecular hydrogen) throughout the simulations. $\mu$ is mean molecular weight, which we adopt a constant value of 2.353, corresponding to the gas of solar metallicity. $k_{\text{B}}$ is the Boltzmann constant, and $m_{\text{H}}$ is the hydrogen atom mass.

The initial values of $\rho$, $P$ and $\pmb{v}$ of gas in the wind tunnel is set as $\rho_{\infty}$, $P_{\infty}$ and $v_{\infty}$, same as the properties of unperturbed disk gas far away from the planet. The Mach number is defined as $\mathcal{M} = v_{\infty}/c_{\infty}$, where the sound speed of the disk is $c_{\infty} = \sqrt{\gamma P_{\infty}/\rho_{\infty}}$.

\subsection{The Model and Numerics} \label{subsec:model}
\subsubsection{The Grid} \label{subsubsec:grid}
The simulations in this work are performed using the open-source high-order Godunov-type hydrodynamics code \texttt{PLUTO} (version 4.1) \citep{Mignone2007}, coupled with a module \texttt{MAKEMAKE} solving radiation transport with FLD approximation \citep{Kuiper2010}. We use the second-order Runge-Kutta (\textit{RK2}) time-stepping scheme, a van-Leer flux limiter in the reconstruction step and a Harten-Lax-van Leer (\textit{hll}) solver to solve the Riemann problem. With \texttt{MAKEMAKE} the FLD equation Eqn. \ref{Eq6} is solved with a generalized minimal residual (GMRES) solver, first developed by \cite{Saad1986}.

We adopt a 2D axially symmetric spherical grid ($r, \theta$) and place the planet object at the origin of the coordinates. The spherical geometry is proven more natural to the problem set-up and yields more accurate solutions to the expected atmosphere structure \citep{Ormel2015a}. The computational domain expands from 1 to 1000 radius of the planet with 538 cells in the $r$ direction and 0 to $\pi$ with 250 cells in the $\theta$ direction. The direction of the pole points to $\theta = 0$. We use logarithmic spacing for the cells in the radial dimension to obtain a higher resolution in the proximity of the planet, where the majority of the proto-atmosphere is located. 

\subsubsection{Boundary and Initial Conditions} \label{subsubsec:bound_init}
The boundary conditions of the computational domain are similar to \cite{Thun2016}. The inner radial boundary of the domain is set as reflective so that no mass or radiation is transported through the planet surface. The outer radial boundary condition is configured according to where the gas flows in and out of the domain. When $0 \leq \theta < \pi/2$, we use the inflow boundary condition to set the ghost cells with the unperturbed gas values. When $\pi/2 \leq \theta \leq \pi$, a zero-gradient condition is implemented instead to ensure the ghost cell values stay the same as the boundary cells so no reflections would occur. We adopt axisymmetric boundary conditions for both boundaries in the polar dimension.

We use the well-known ``Minimum Mass Solar Nebula'' (MMSN) model \citep{Weidenschilling1977, Hayashi1981, Thommes2006} to provide the approximate initial values for the gas properties in the unperturbed disk. 
In our canonical simulation, we set up a planetary object (M$_{\text{pl}}$ = 1 M$_{\oplus}$, R$_{\text{pl}}$ = 1 R$_{\oplus}$) embedded in the homogeneous disk gas (T$_{\infty}$ = 300 K, $\rho_{\infty}$ = 10$^{-9}$~g cm$^{-3}$) located approximately at $a =$ 1 au from the Sun. 
The relative velocity of the planet to the gas $v_{\infty}$ is set as 4.8 km s$^{-1}$ (or $\mathcal{M} = 3.93$).
Planets with orbital eccentricities between 0.17 to 0.35 are able to reach such a velocity at certain positions in their orbits. 
Adopting values provided from different disk models would slightly differ the background density value for the location of the planet. For example, the updated MMSN model \citep{Desch2007} would increase the nebular density at 1 au by 1 to 1.5 orders of magnitude, while the ``Minimum Mass Extra-solar Nebula'' (MMEN, \citealt{Chiang2013, Lee2014}) would increase it by 0.5 to 1 orders of magnitude. These variations are within the range of background gas density considered in our parameter study (\S~\ref{subsubsec:rho_bkg}). 

More accurately, the density profile of the background gas medium follows an approximated vertical hydrostatic stratification structure for protoplanetary disks (e.g. \citealt{Armitage2010}):
\begin{equation}
    \rho_{\infty}(z) = \rho_{\infty}\ \text{exp}\left(-\frac{z^2}{2H^2}\right)
\end{equation}
where $z$ is the vertical distance from the midplane, $H = \sqrt{\mathcal{R}_{\text{H}_2}Ta^3/(GM_{\odot})}$ is the disk gas pressure scale height ($\mathcal{R}_{\text{H}_2}$ is the specific gas constant for H$_2$, $M_{\odot}$ is the solar mass). For simplicity, we do not take into account the vertical density profile but adopt the midplane gas density $\rho_{\infty}$ as the background value. This is because we are concentrated on relatively small planets or planetary embryos (e.g. sub-Earth- or Earth-size objects). The radii of these objects (including the proto-atmospheres) are at least four orders of magnitude smaller than $H$ in our simulations.

The initial set-up is slightly different in our orbit-dependent simulation (see \S~\ref{subsec:list}) as we start the simulation when the planet is at its perihelion. The initial values of gas properties ($\rho_{\infty}, T_{\infty}, P_{\infty}, v_{\infty}$) are adjusted to match the corresponding disk environment for an orbit with the semimajor axis $a =$ 1 au and eccentricity $e =$ 0.2. 

\subsubsection{Opacities} \label{subsubsec:opac}
As mentioned above, the homogeneous background gas is assumed to have the same composition as the solar nebula. 
Data for gas opacity is adopted from \cite{Malygin2014}. Opacities for lower temperatures (i.e. $T_{\text{gas}} <$ 700 K) are extrapolated from the higher temperature data (see \cite{Marleau2019} for details).
We include dust in all simulations with a dust-to-gas mass ratio $\sim$ 0.01, assuming ``normal'' silicates with Fe/(Mg+Fe)=0.3, as defined in \cite{Semenov2003}. Dust grains are simplified as homogeneous spherical particles, and their opacities are taken from corresponding tables in \cite{Semenov2003} without the built-in dust evaporation calculations. Instead, evaporation of refractory component in the dust-to-gas transition temperature region (1400 - 1600 K) is modeled separately -- we adopt a smooth transition curve to lower temperatures with arctan functions and steep linear decrease to higher temperatures.

\subsection{List of Simulations} \label{subsec:list}
The main goal of this work is to explore how an eccentric orbit affects the dynamics and structure of the planetary proto-atmosphere. 
To achieve this goal and answer the four science questions in \S~\ref{sec:intro} we carry out a list of simulations with different initial conditions, which can be divided into the ``snapshot'' simulations and the orbit-dependent simulation (see Table \ref{tab:list_sim}).

For the ``snapshot'' simulations, we set the relative velocity between the planet and disk gas ($v_{\infty}$, or represented by the Mach number $\mathcal{M}$), the planet mass (M$_{\text{pl}}$) and gas density ($\rho_{\infty}$) as free input parameters to specify the disk-planet environment. The gas pressure is adjusted to match the gas temperature at 300 K. The simulations are run until they reach steady/quasi-steady states so that the solutions represent ``snapshots'' of the proto-atmospheres in the corresponding environments. 
For the orbit-dependent simulation, we use the planet mass ($M_{\text{pl}}$), the orbital semimajor axis ($a$) and orbital eccentricity ($e$) as inputs to determine the specific eccentric orbit for the planet.  Other parameters such as gas density, pressure, temperature, and relative velocity are determined upon the time-dependent star-planet distance, which in turn, is dependent on these input orbital parameters.
It is, therefore, a time-dependent simulation that we read outputs regularly to track the proto-atmosphere evolution throughout the eccentric orbit. 
Given that these simulations are significantly more expensive in terms of computation hours, we do not perform a parameter study involving full-orbit simulations in this work.

\begin{figure*}[ht!]
\centering
\includegraphics[scale=0.45]{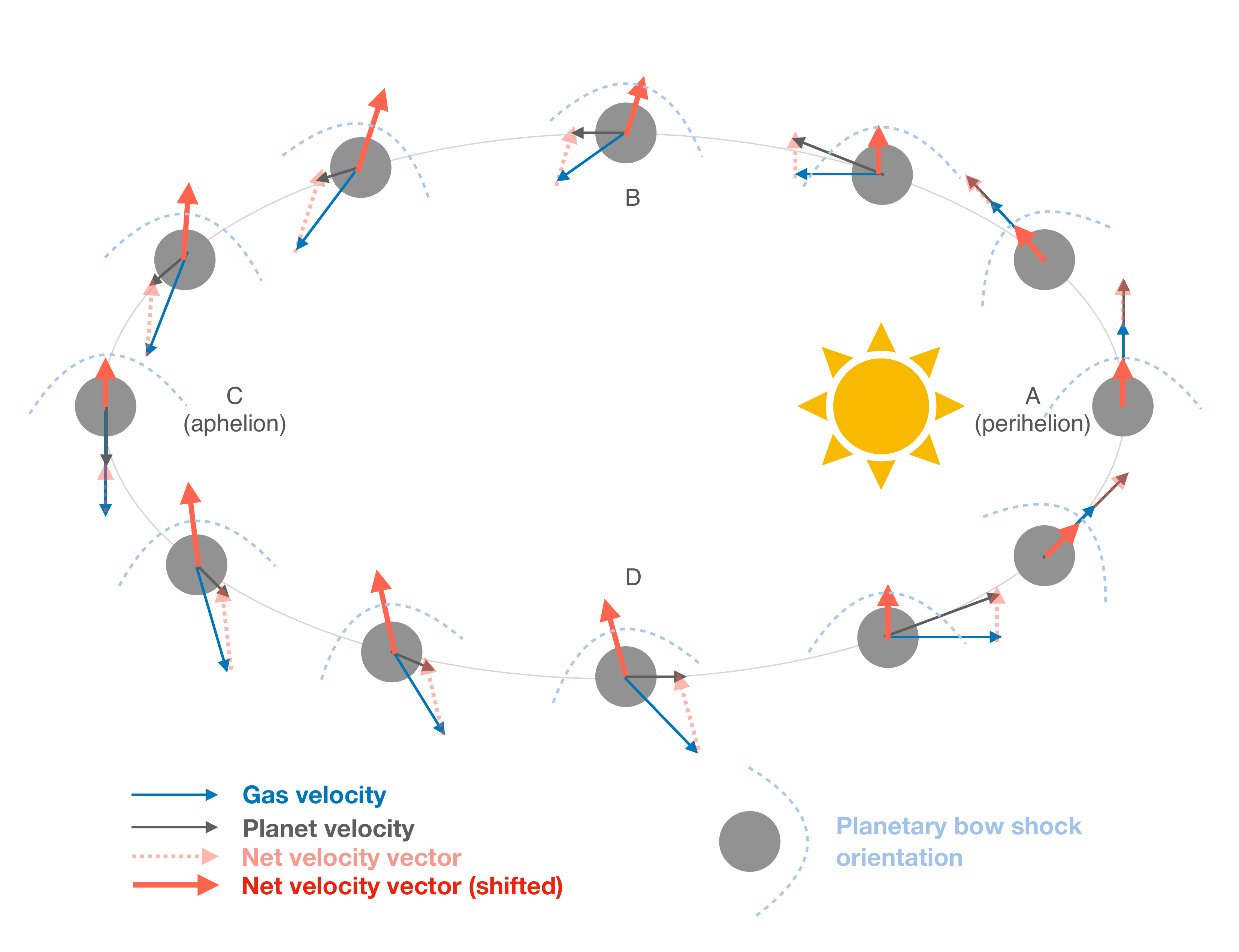}
\caption{Schematic of a planet traveling supersonically on an eccentric orbit. The dark gray arrows represent the planet's orbital velocity ($\pmb{\upsilon_{\text{pl}}}$), tangential to the elliptical orbit. The blue arrows denote the disk gas velocity ($\pmb{\upsilon_{\text{gas}}}$), which are always perpendicular to the direction of the star. The relative velocity between the planet and the gas are the vector difference of the dark gray and blue arrows (the dotted arrows in light red, also shifted to the center of the planet as solid red arrows). 
Note that the schematic has exaggerated the orbital eccentricity and the planet size. The absolute lengths of all arrows in the figure have little meaning. Both $\textbar \pmb{\upsilon_{\text{pl}}}\textbar$ and and $\textbar \pmb{\upsilon_{\text{gas}}}\textbar$ decrease as the planet gets further away from star, though $\textbar \pmb{\upsilon_{\text{pl}}}\textbar > \textbar \pmb{\upsilon_{\text{gas}}}\textbar$ from location A to B, or D to A, and $\textbar \pmb{\upsilon_{\text{pl}}}\textbar < \textbar \pmb{\upsilon_{\text{gas}}}\textbar$ from location B to D.
The red vectors also represent the orientation of planetary shock fronts (symbolized as light blue dotted curves). Counter-intuitively, the orientation of bow shocks does not change significantly throughout the orbit.
Given that the orbital time scale is much larger than the dynamical time scale of gas accretion, the change of bow shock orientation is trivial in the local-framed simulations.
\label{fig:Schem_orbit}}
\end{figure*}

In reality, the orientation of the planetary bow shock is also changing as the planet moves to different locations on the orbit. It is directly determined by the vector difference between the planet orbital velocity ($\pmb{\upsilon_{\text{pl}}}$) and the disk gas velocity ($\pmb{\upsilon_{\text{gas}}}$) at each location (see Fig.~\ref{fig:Schem_orbit}). Both quantities decrease in magnitude with a larger distance from the star. At perihelion, the planet travels faster than the gas and generates a shock front aligned with $\pmb{\upsilon_{\text{pl}}}$. At aphelion, the gas in sub-Keplerian motion is faster and the shock front is in the opposite direction as $\pmb{\upsilon_{\text{pl}}}$. As shown in Fig.~\ref{fig:Schem_orbit}, the change of bow shock orientation throughout the orbit is limited (max. $\sim$ 20$^{\circ}$ for an orbit $e = 0.2$). Because the timescale on which the orientation changes is much longer than the dynamical timescale of gas accretion, the variation of bow shock direction is therefore negligible in the local-framed hydrodynamic simulations.

\begin{table*}[!ht]
\centering
\caption{List of Hydrodynamic Simulations}
\label{tab:list_sim}
\begin{threeparttable}
\begin{tabular*}{0.75\textwidth}{ccccc}
\hline
\hline
\multicolumn{5}{c}{\textbf{``Snapshot'' Simulations}} \\
Name        &  $v_{\infty}$ (km s$^{-1})$ / $\mathcal{M}$    &     M$_{\text{pl}}$ (M$_{\oplus}$) &      $\rho_{\infty}$ (g cm$^{-3}$)   &   Comments          \\ \hline
\textbf{\texttt{Canonical}}       &    \textbf{4.8} / \textbf{3.93}    &    \textbf{1.0}          &   \textbf{10$^{-9}$}   &   \makecell{\textbf{The canonical case}}        \\
\texttt{VEL3.2}           &    3.2 / 2.64     &    1.0          &   10$^{-9}$   &   \multirow{8}{*}{\makecell{Sensitivity study \\ on Mach number/ \\relative velocity}}        \\
\texttt{VEL3.7}          &   3.7 / 3.02   &    1.0            &    10$^{-9}$   &      \\ 
\texttt{VEL4.2}          &   4.2 / 3.47   &    1.0            &    10$^{-9}$   &      \\
\texttt{VEL5.3}          &   5.3 / 4.38   &    1.0            &    10$^{-9}$   &      \\
\texttt{VEL5.9}          &   5.9 / 4.84   &    1.0            &    10$^{-9}$   &      \\
\texttt{VEL6.4}          &   6.4 / 5.29   &    1.0            &    10$^{-9}$   &      \\
\texttt{VEL7.0}          &   7.0 / 5.75   &    1.0            &    10$^{-9}$   &      \\
\texttt{VEL7.5}           &   7.5 / 6.2    &    1.0            &    10$^{-9}$   &      \\ \hline
\texttt{PM0.1}          &   4.8 / 3.93    &    0.1            &    10$^{-9}$   &   \multirow{6}{*}{\makecell{Sensitivity study \\ on planet mass}}     \\
\texttt{PM0.3}          &   4.8 / 3.93    &    0.3            &    10$^{-9}$   &      \\
\texttt{PM0.5}          &   4.8 / 3.93    &    0.5            &    10$^{-9}$   &      \\
\texttt{PM0.7}          &   4.8 / 3.93    &    0.7            &    10$^{-9}$   &      \\
\texttt{PM2.0}          &   4.8 / 3.93    &    2.0            &    10$^{-9}$   &      \\
\texttt{PM3.0}          &   4.8 / 3.93    &    3.0            &    10$^{-9}$   &      \\ \hline
\texttt{DEN-7}          &   4.8 / 3.93    &    1.0            &    10$^{-7}$   &    \multirow{4}{*}{\makecell{Sensitivity study \\ on gas density}}    \\
\texttt{DEN-8}          &   4.8 / 3.93    &    1.0            &    10$^{-8}$   &    \\
\texttt{DEN-10}          &   4.8 / 3.93    &    1.0            &    10$^{-10}$   &   \\
                                       \hline
\multicolumn{5}{c}{\textbf{Orbit-dependent Simulation}} \\
Name     &    M$_{\text{pl}}$ (M$_{\oplus}$)     &   a (au)  &  e        &   Comments    \\  \hline
\texttt{ORB}   &   1.0    &   1.0   &   0.2    & \makecell{Time-dependent \\ simulation to track \\ orbital evolution}  \\
\hline

\end{tabular*}%

\end{threeparttable}
\end{table*}

\section{Results} \label{sec:results}
We present the general 2D flow structure of the bow shock and the proto-atmosphere around the planet in \S~\ref{subsec:canonical}. We conduct a parameter study on the ``snapshot'' simulations and show how the solutions change with the planet--disk environments in \S~\ref{subsec:sensitive}. Results for the orbit-dependent case are shown in \S~\ref{subsec:orbit}.

\subsection{The Canonical Case} \label{subsec:canonical}
\subsubsection{The flow structure} \label{subsubsec:flow}
When given enough time, all the ``snapshot'' simulations are able to reach steady states where the atmospheres reach their final mass and the accretion rates approach zero. Note that we do not take into account the atmosphere contraction occurring in the Kelvin--Helmholtz timescale, which is too long to be modeled in radiation hydrodynamic simulations.
Depending on the specific set-up, the time required to reach the steady state varies. To avoid extensive computational time for some simulations to reach steady states, they are run only until they reach quasi-steady states\footnote{We define that the system reaches a quasi-steady state where the atmosphere mass growth becomes trivial -- the time to grow another 0.1\% of the total mass is much longer than the time to obtain the majority (90\%) of the final mass.}. The time required for the system to reach a quasi-steady state is also the dynamical timescale. For the canonical case, the timescale is $\sim 6\times 10^6$~s. We then fit the temporal growth of the atmosphere mass with an asymptotic function to estimate the approximate final mass (see Appendix \ref{append:steady} for a detailed description).


\begin{figure*}[ht!] \label{fig:rho_Tgas}
\centering
\gridline{\fig{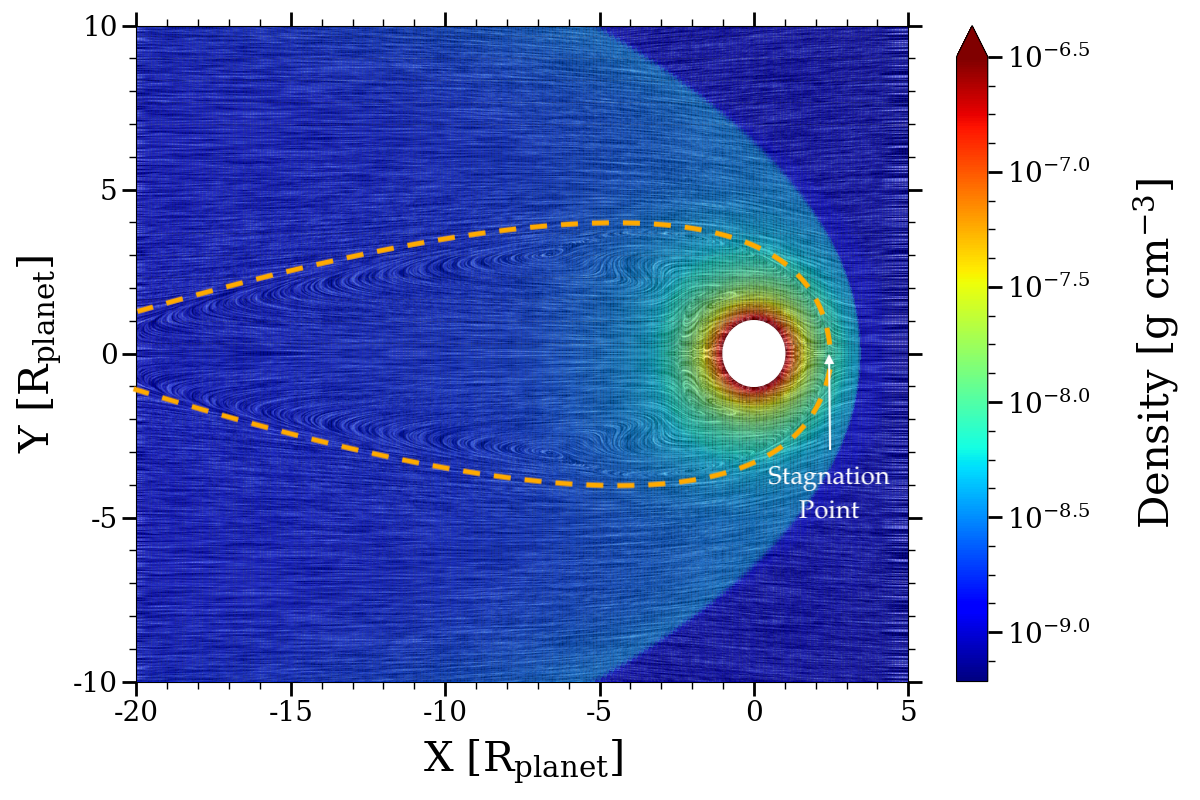}{0.5\textwidth}{(a)}
         \fig{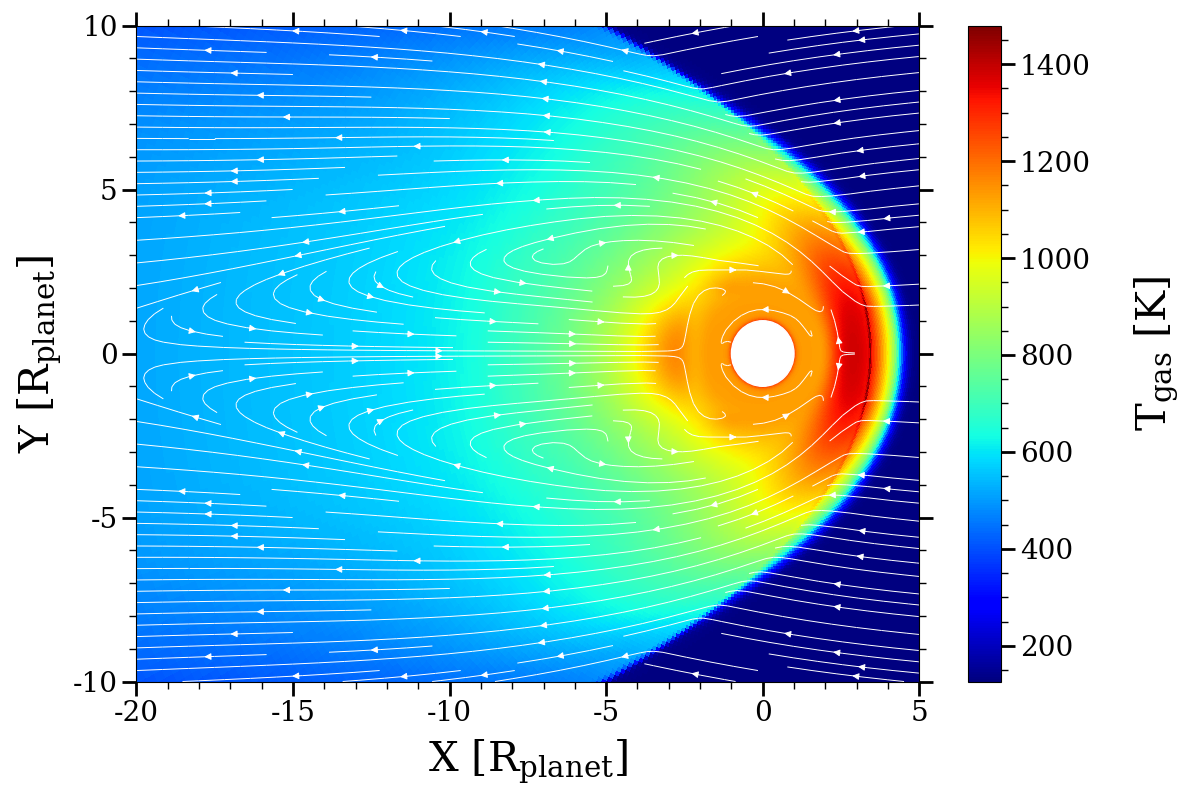}{0.5\textwidth}{(b)}
          }
\gridline{\fig{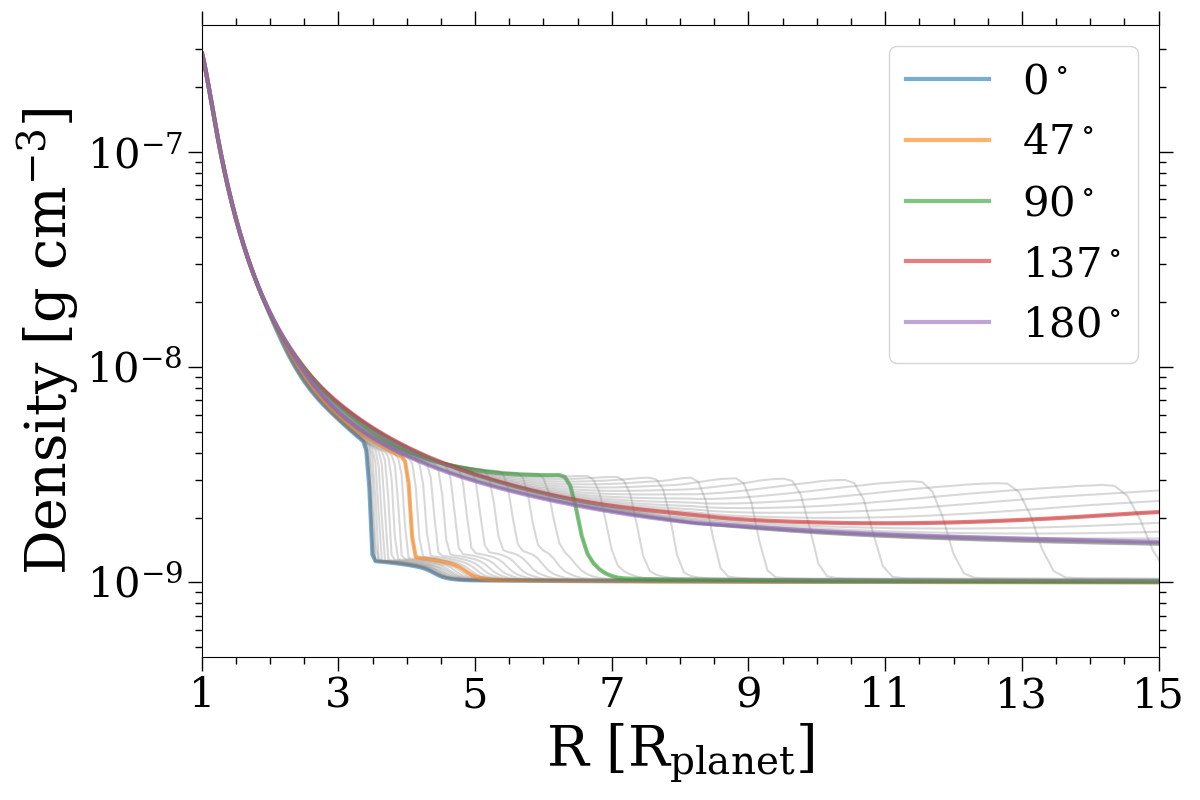}{0.4\textwidth}{(c)}
         \fig{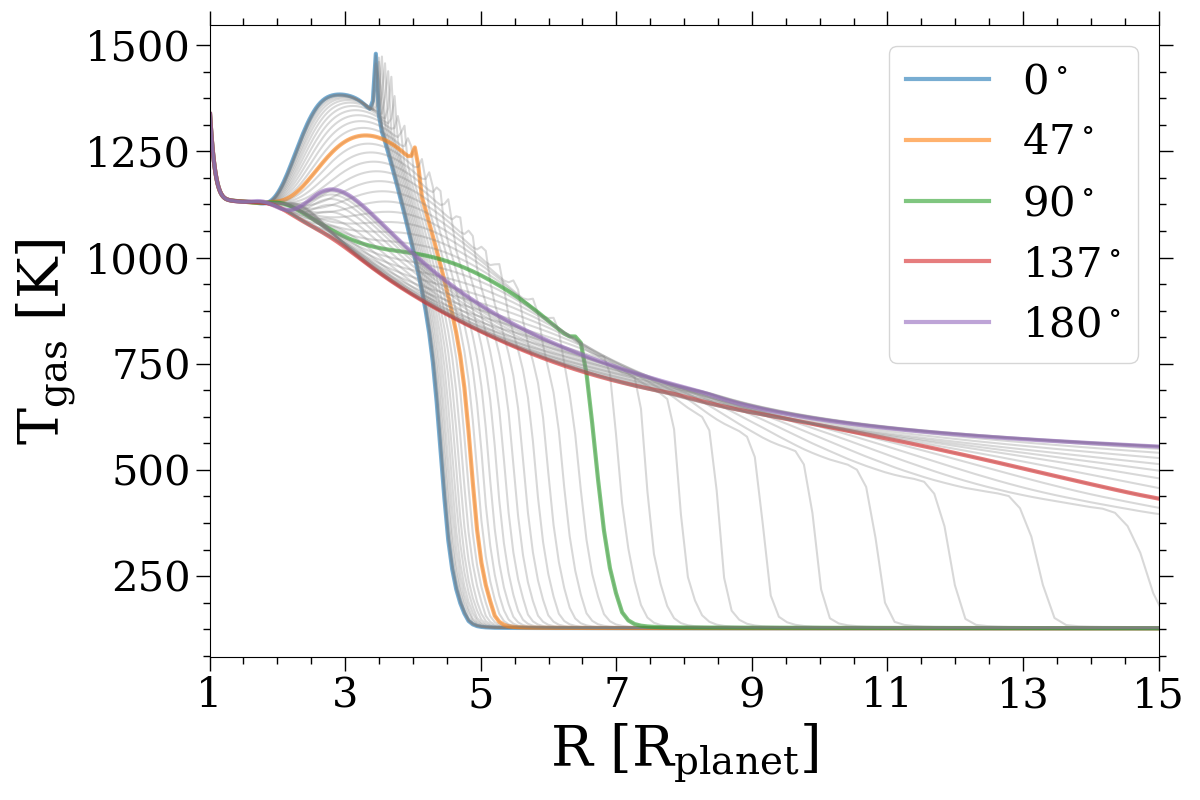}{0.4\textwidth}{(d)}
          }
\caption{Hydrodynamical solutions for the canonical case in quasi-steady state (M$_{\text{pl}}$ = 1.0 M$_{\oplus}$, $\rho_{\infty}$ = 10$^{-9}$ g cm$^{-3}$, $v_{\infty}$ = 4.8 km s$^{-1}$). a: Gas density and the flow pattern illustrated with Linear Integral Convolution (LIC). A hydrostatic region is formed around the planet in the bow shock. Depending on where the gas enters the bow shock, it either gets deflected and directly flows to downstream or flow back to the planet, forming multiple stable vortices inside the ``egg-shape'' flow region (also one of the definitions for a proto-atmosphere, enclosed by the orange dashed line). In general, the gas entering the bow shock within $+/-$ 2 R$_{\text{pl}}$ flows inside the ``egg-shape'' region. The bow shock stagnation point locates in front of the planet on the yellow dashed line. b: Gas temperature and streamlines. Note that the temperature gradient is generally smooth everywhere except at the shock front (the sharp jumps at X = $-9$ and X = $-4$ are merely caused by the color transition). c: 1D profiles of gas density at different angles drawn as gray lines ($\theta$, with $\theta = 0$ corresponding to the positive $x$ axis). Solutions along $\theta \sim 0, \pi/4, \pi/2, 3\pi/4, \pi$ are highlighted with colors. d: 1D temperature profiles with different angles drawn as gray lines. The temperature solutions show a typical supercritical shock. The gas is heated beyond the shock front by the radiation from the post-shock region and has a temperature spike at the density shock front. See \S~\ref{subsubsec:flow} for more details.}
\end{figure*}

In Fig.~\ref{fig:rho_Tgas} we present the hydrodynamical solutions (gas density and temperature) of the canonical simulation in a quasi-steady state (M$_{\text{pl}}$ = 1.0 M$_{\oplus}$, $\rho_{\infty}$ = 10$^{-9}$ g cm$^{-3}$, $v_{\infty}$ = 4.8 km s$^{-1}$). Panel a and b show the 2D solutions while c and d show the 1D profiles along different angles.

A strong bow shock is formed in front of the planet with the shock front locates at 3.6 $R_{\text{pl}}$ (planet radii). The ram pressure from the nebular gas and the bow shock prevent a large amount of gas to be accreted to the planet on the side directly facing the shock front. But the planet still retains a proto-atmosphere from the post-shock gas that is largely hydrostatic in a steady/quasi-steady state. 
The proto-atmosphere is largely pressure-supported.
The nebular gas is heated at the shock front due to strong compression and pre-heated beyond the shock front by the radiation emitting from the hot post-shock gas.
This is a type of radiation shock called supercritical shock \citep{Sincell1999} featuring a temperature spike at the density shock front (Fig.~\ref{fig:rho_Tgas}d).
The stronger the shock is, the higher the temperature spike can be and the larger the pre-heated region beyond the shock front is.
Fig.~\ref{fig:rho_Tgas}d also shows that the inner proto-atmosphere is mainly heated by the accretional energy, and temperature of the outer layer (two to three planet radii) is mildly raised by the bow shock compression. 
There is also a small high-temperature region behind the planet caused by the converging flows from vortices.
We illustrate the gas flow with the Linear Integral Convolution (LIC, Fig.~\ref{fig:rho_Tgas}a) and streamlines (Fig.~\ref{fig:rho_Tgas}b). 
Both illustrations show a clear boundary between two types of flow -- gas entering the bow shock with larger impact parameters directly flow past the planet, and gas with relatively small impact parameters flowing back towards the planet in the downstream region, forming multiple stable vortices (``egg-shape'' region, enclosed by the yellow dashed line in Fig.~\ref{fig:rho_Tgas}a). 

The ``egg-shape'' flow region resides in the post-shock subsonic area, with its boundary in front of the planet coincided with the stagnation point of the bow shock. 
Interestingly, the proto-atmosphere is accreted not directly through the gas encountering the front of the planet, but the gas flowing back towards the planet in the ``egg-shape'' pattern. 
Such patterns are clearly distinctive from those of gas accretion on planets with circular orbits, where gas is accumulated from all directions (e.g. \citealt{Ormel2015a, Ormel2015b, Cimerman2017}). 
The flow structure becomes stable soon after the bow shock is established.

\begin{figure*}[ht!]
\centering
\includegraphics[scale=0.24]{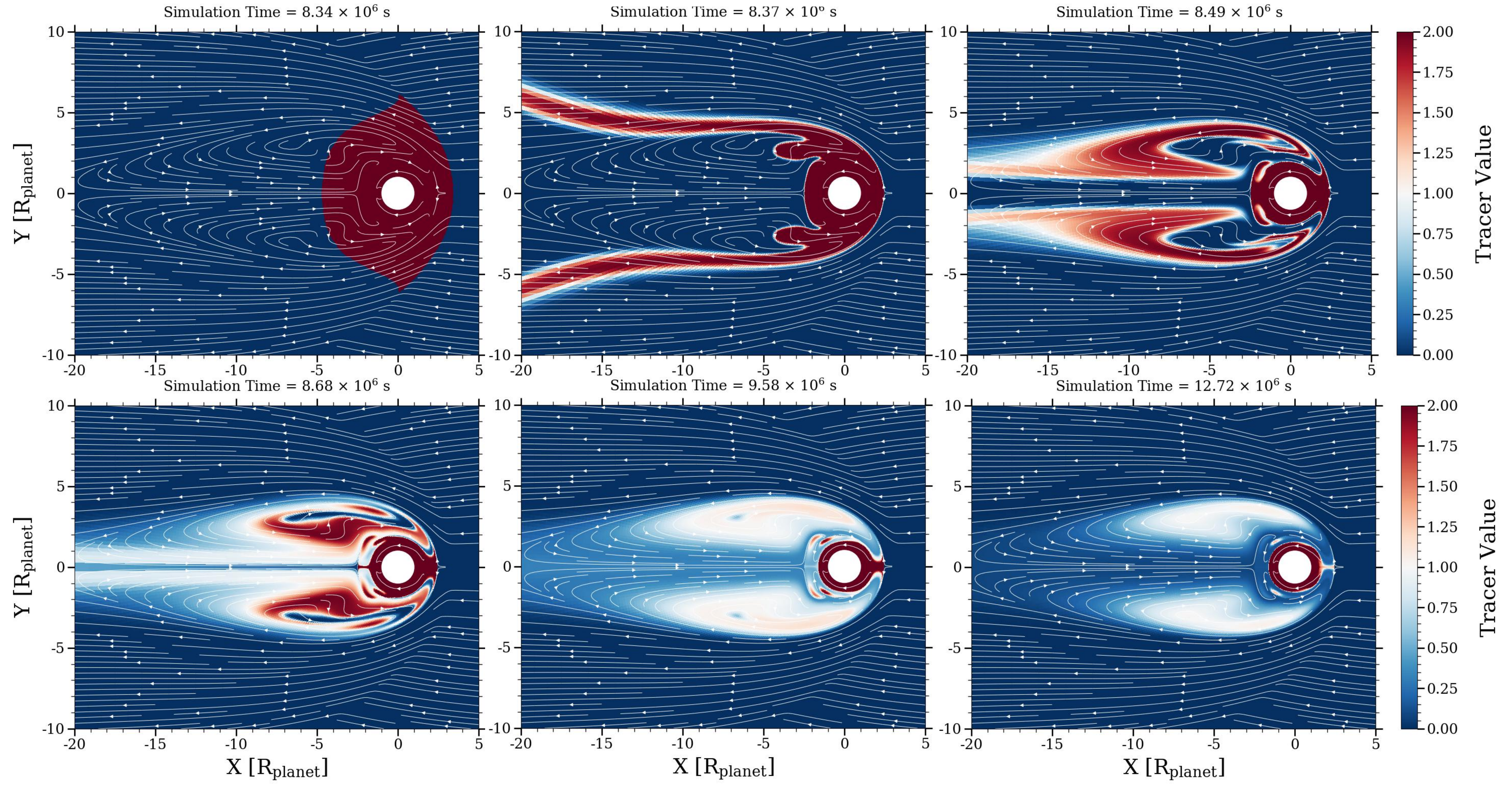}
\caption{The movements of gas in the proto-atmosphere tracked with a tracer (initial value = 2) after the simulation reaches a quasi-steady state reveal the recycling of disk gas behind the planet. We set the initial region with dye as the part of atmosphere enclosed by the 10$^{-8.5}$~g cm$^{-3}$ density contour (one of the definitions of proto-atmospheres, see \S~\ref{subsubsec:atm_mass}). The outermost layer of gas outside of the ``egg-shape'' region directly flows out the computational domain, while the majority of gas circulates in vortices behind and around the planet. A portion of the gas stays bound around the planet, but the rest would flow out of the ``egg-shape'' region and get replaced by new incoming disk gas. \label{fig:tracer}}
\end{figure*}

As the simulation reaches a quasi-steady state, we dye the proto-atmosphere region with a tracer to track the gas movement over $\sim 5 \times 10^6$~s, a time long enough to see the gas replacement inside the ``egg-shape'' region (Fig.~\ref{fig:tracer}). 
Although the tracer diffuses over time, it is clear that the majority of the dyed gas flow back to the planet inside the “egg-shape” region. A portion stays bound close to the planet, while the rest eventually flows out of vortices. These vortices then get replenished by the next batch of incoming gas. The gas inside the ``egg-shape'' region is constantly recycled and exchanging with the disk gas, with an estimated recycling timescale between 5$\times$10$^5 \sim 5\times$10$^6$~s\footnote{The timescale estimation is based on whether the majority of gas inside the ``egg-shape'' region is replaced. There is a range for this timescale because of the uncertainty caused by the diffusion of the tracer.}, about half of the dynamical timescale.

\begin{figure}[ht!]
\centering
\includegraphics[scale=0.28]{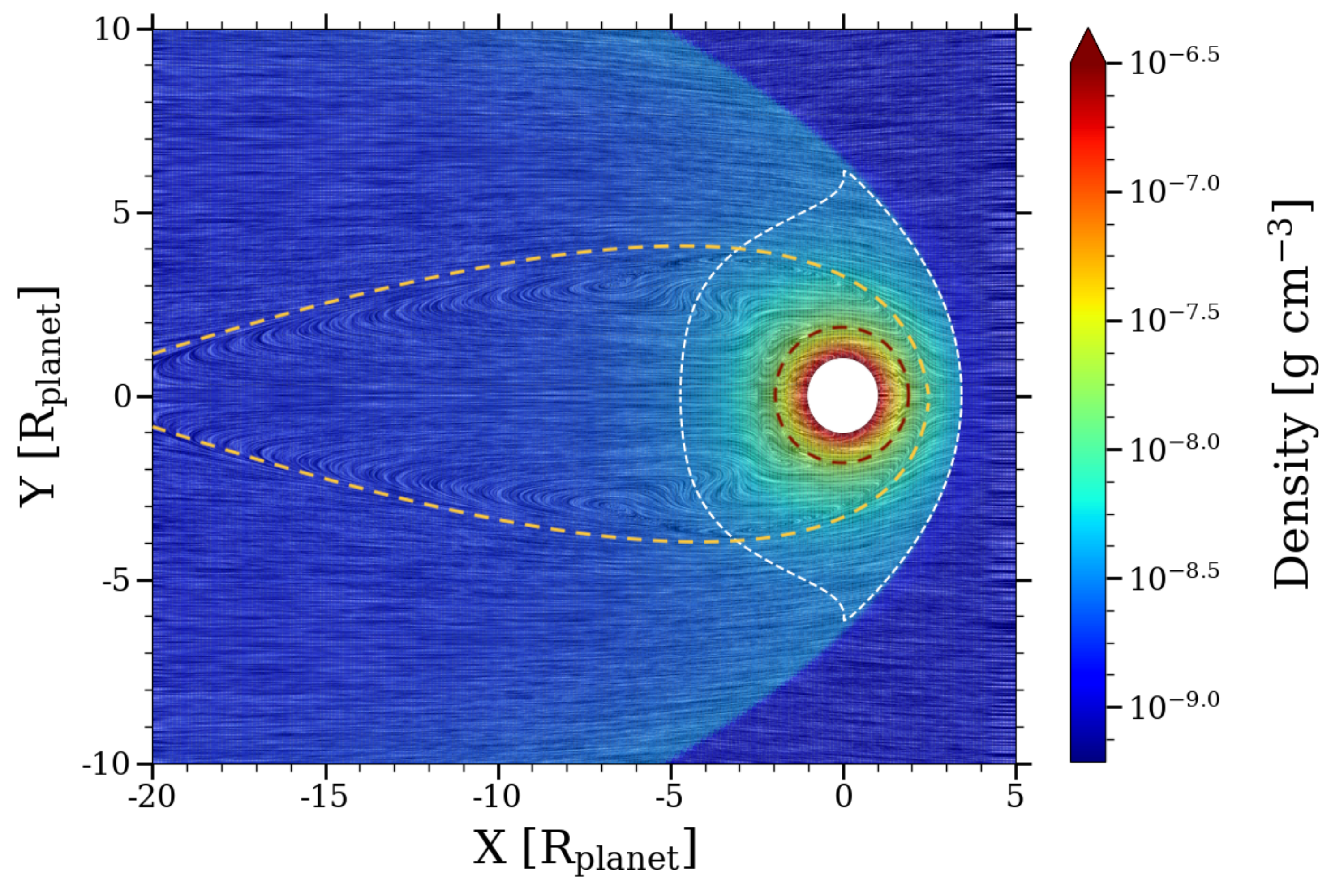}
\caption{The three definitions of proto-atmospheres on eccentric planets. The white dashed line is the 10$^{-8.5}$~g cm$^{-3}$ density contour, the yellow dashed line denotes the boundary of the ``egg-shape'' region and the red dashed line encloses the gas that is bound to the planet. The gas recycling region is between the red and yellow dashed lines. \label{fig:define}}
\end{figure}

\subsubsection{Definitions of the Atmosphere} \label{subsubsec:atm_mass}
Defining the boundary of the proto-atmosphere of an eccentric planet is far less straightforward compared to the cases where planets are on circular orbits. The commonly used definitions for symmetric atmospheres such as Hill radius and Bondi radius are no longer applicable in the supersonic environment.
Like those around planets on circular orbits, the proto-atmosphere is always connected to and actively interacting with the disk gas (\S~\ref{subsubsec:flow}). 

For the purpose of calculating the atmosphere mass, we provide three different ways to define a boundary for the proto-atmosphere based on the flow structures resolved by the hydrodynamic simulation --
\begin{enumerate}
    \item A gas density contour around the planet that encloses the hydrostatic atmosphere and not too distorted in shape because of the bow shock. We found that the 10$^{-8.5}$~g cm$^{-3}$ contour is most suitable when the background gas density $\rho_{\infty} = 10^{-9}$~g cm$^{-3}$. The mass enclosed by this boundary is 1.40 $\times$ 10$^{21}$~g.
    
    \item The boundary of the ``egg-shape'' region where the gas recycles in and out. There is 1.43 $\times$ 10$^{21}$~g of gas in this region when the system reaches a steady/quasi-steady state. 
    
    \item The boundary of the gas that stays gravitationally bound to the planet, according to Fig.~\ref{fig:tracer}. The mass of this part of gas is about 4.31$\times$ 10$^{20}$~g.
\end{enumerate}{}

Fig.~\ref{fig:define} demonstrates the different atmosphere boundaries by these definitions with dashed lines in different colors. As we will see in \S~\ref{subsec:sensitive}, the gas enclosed by the 10$^{-8.5}$~g cm$^{-3}$ contour line and the gas in the ``egg-shape'' region are of similar mass, both about half an order of magnitude higher than the bound mass. By subtracting the bound mass from the mass in the ``egg-shape'' region, we obtain the amount of recycling gas behind the planet to be 9.98$\times$ 10$^{20}$~g.

We would like to point out that the atmospheric pressure on the planet surface is a probably better metric of how much gas is retained.
Because in the current context, the calculated atmosphere mass is relatively subjective --- it depends on which boundary is used. In comparison, the surface pressure does not rely on a specific definition. The larger the surface pressure, the thicker the atmosphere. In cases where the planet surface is partially or completely molten, the surface pressure directly determines how much volatiles can be dissolved in the magma ocean (e.g. \citealt{Wu2018, Sharp2019}).
The average surface pressure for the canonical case is 4 $\times$ 10$^{-2}$ bar. We use both surface pressure and the atmosphere mass to measure the amount of the proto-atmosphere throughout the study.

    
    

\subsection{Study of Sensitivity to Parameters} \label{subsec:sensitive}
Does the proto-atmosphere of an eccentric planet always have a similar flow structure as seen in the canonical simulation? How does it change in response to different planet--disk environments? To answer these questions, we vary the three input parameters: the relative velocity between the planet and the gas ($v_{\infty}$), the planet mass ($M_{\text{pl}}$) and background gas density ($\rho_{\infty}$) separately and carry out the corresponding ``snapshot'' simulations (see Table \ref{tab:list_sim}).

\begin{figure*}[ht!]
\centering
\includegraphics[scale=0.24]{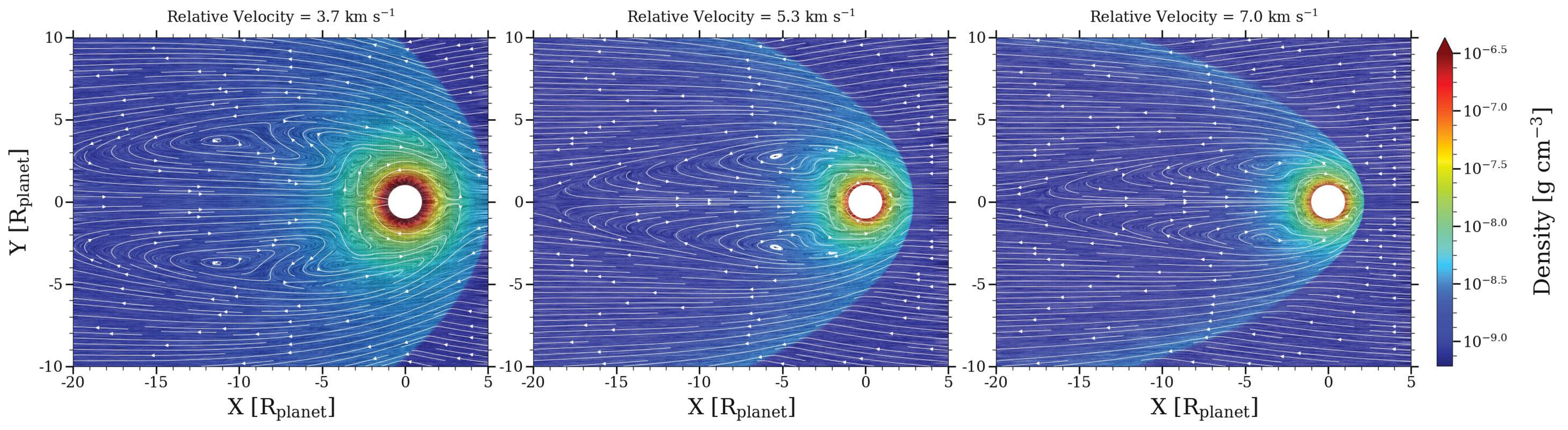}
\caption{Gas density and the flow structure around the planet with various relative velocity between the planet and the disk gas (3.7, 5.3 and 7.0 km s$^{-1}$ respectively), illustrated with LIC and streamlines. In general, the hydrodynamics remain the same as the canonical case (Fig.\ref{fig:rho_Tgas}). The planet is able to hold a thicker and more massive atmosphere at a low velocity relative to the disk gas. The ``egg-shape'' flow pattern persists in all simulations performed, though the area reduces as the bow shock becomes stronger in high relative velocity cases. \label{fig:param_vel2D}}
\end{figure*}

\begin{figure}[ht!]
\centering
\includegraphics[scale=0.28]{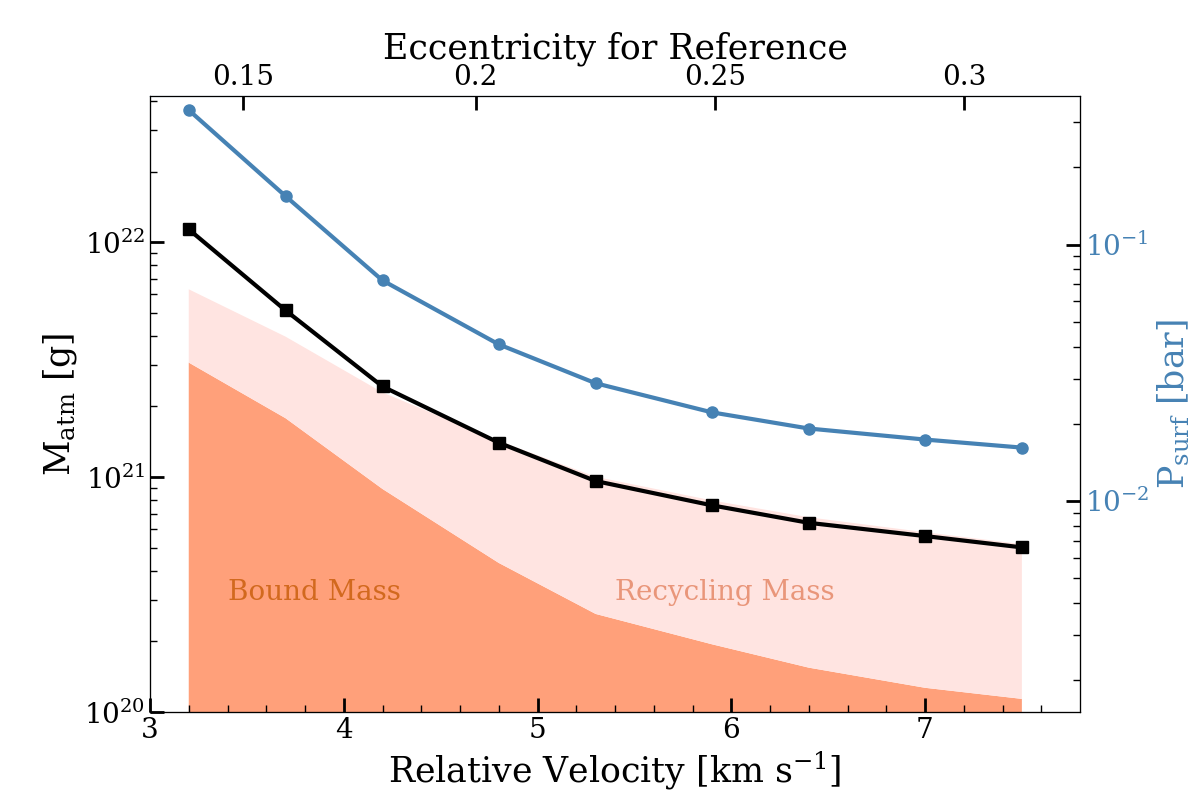}
\caption{The impact from the relative velocity between the planet and the disk gas ($v_{\infty}$) on the atmosphere mass (black markers and line, defined by the density contour 10$^{-8.5}$~g cm$^{-3}$) and averaged planet surface pressure (blue markers and line). The mass of the bound gas is shaded with deep orange color, while the mass of the recycling gas is colored as light orange. The two sums up to be the gas mass inside the ``egg-shape'' region behind the planet, which is very close to the atmospheric mass enclosed by the density contour. 
The lowest and highest relative velocities considered here are the average velocities experienced by planets on eccentric orbits with e  $\sim$ 0.15 up to e $\sim$ 0.3 (top x axis), although a planet does not experience a single relative velocity over its orbit (Fig.~\ref{fig:relative_vel})}.\label{fig:Param_Mach}
\end{figure}

\subsubsection{The Effect of Relative Velocity} \label{subsubsec:Mach}
The relative velocity between the planet and the disk gas directly determines the bow shock strength around the planet when it exceeds the sound speed ($\sim$ 1.2 km s$^{-1}$ for the assumed background disk gas). In the parameter study, we vary the relative velocity between 3.2 and 7.5 km s$^{-1}$, corresponding to planets with orbital eccentricities between 0.1 and 0.5.

The hydrodynamical solutions of these simulations reveal that the flow structures around the planet are similar to the canonical case. The ``egg-shape'' region forms behind the planet with part of the gas bound to the planet and the rest exchanging with the disk, regardless of how fast the planet is traveling through the gas (Fig.~\ref{fig:param_vel2D}). As the relative velocity gets larger, the bow shock opening angle becomes narrower. The larger ram pressure coming from the nebular gas suppresses the atmosphere accretion further and reduces the stand-off distance of the shock front. As a result, the planet keeps a smaller atmosphere with an even smaller portion of the bound part.
Fig.~\ref{fig:Param_Mach} shows how the atmosphere mass and planet surface pressure change with the variation of the relative velocity. The atmospheric properties show most dramatic changes when the planet is traveling relatively slow in the disk gas (but still supersonic). In the cases of high relative velocity, the ``egg-shape'' region is dominated by the recycling movements of gas.

\begin{figure*}[ht!]
\centering
\includegraphics[scale=0.24]{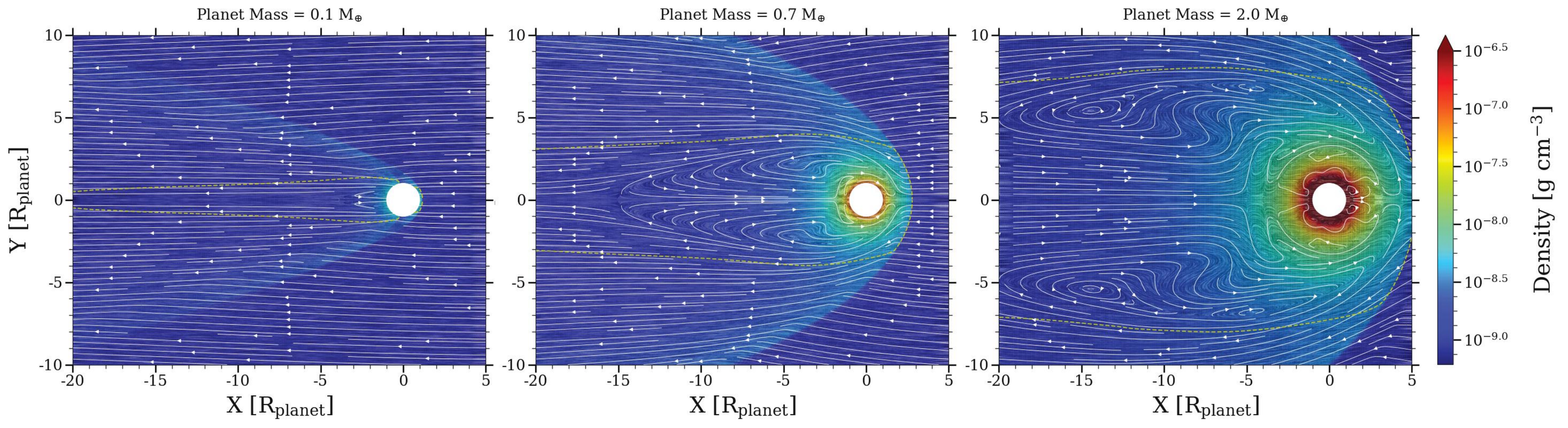}
\caption{Gas density and the flow structure around the planet with various planet mass (0.1, 0.7 and 2.0 M$_{\oplus}$ respectively), illustrated with LIC and streamlines. The planet radii have been adjusted accordingly following the mass-radius relationship in \citep{Unterborn2018, Dressing2015}. The yellow dashed lines show where the gas velocity reduces from supersonic to subsonic. In simulations when the planet mass $\geq$ 0.3 M$_{\oplus}$, the hydrodynamics remain the same as the canonical case (Fig.~\ref{fig:rho_Tgas}). In the case of 0.1 M$_{\oplus}$, the planet cannot hold a regular atmosphere. With larger gravity, the planet pushes the shock front further and retains a more massive atmosphere. The ``egg-shape'' flow pattern also expands and more complicated vortices are developed inside the region. \label{fig:param_pm2D}}
\end{figure*}

\begin{figure}[ht!]
\centering
\includegraphics[scale=0.28]{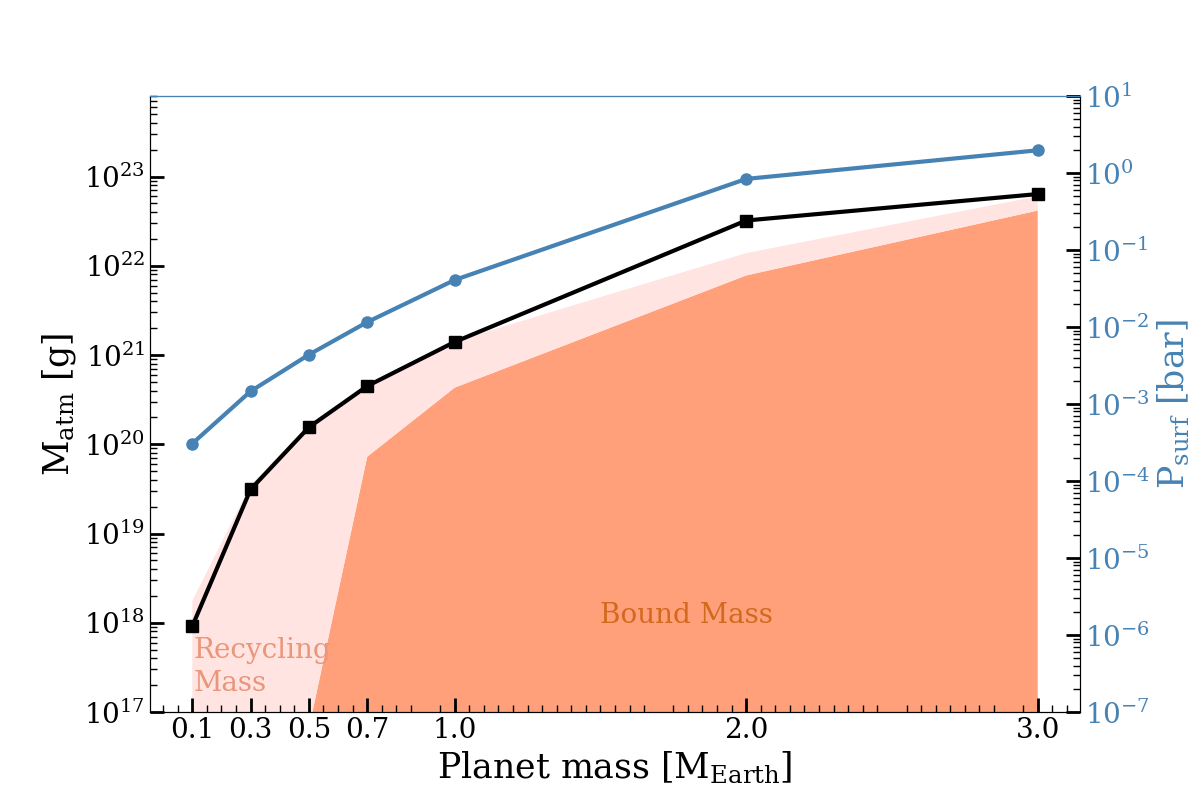}
\caption{Similar to Fig.~\ref{fig:Param_Mach}. The impact from the planet mass ($M_{\text{pl}}$) on the atmosphere mass and averaged planet surface pressure. Both increase with a larger planet mass. \label{fig:Param_pm}}
\end{figure}

\subsubsection{The Effect of Planet Mass} \label{subsubsec:M_pl}
The planet mass directly affects the amount of atmosphere that can be gravitationally accreted onto the planet. We vary the planet mass from 0.1 M$_{\oplus}$ to 3 M$_{\oplus}$, covering a range of terrestrial planets from the sub-Earth to super-Earth regime (Table \ref{tab:list_sim}).
We assume that all simulated planets have Earth-like composition and interior structures (e.g. similar element ratios and temperature structures). Therefore, we calculate the planet radius as a function of planet mass using the \texttt{ExoPlex} mass-radius software package \citep{Unterborn2018} for planets $\leq$ 0.7 M$_{\oplus}$ and mass-radius relationship described in \cite{Dressing2015} for planets $>$ 0.7 M$_{\oplus}$.

Fig.~\ref{fig:param_pm2D} shows the hydrodynamical solutions for planets with 0.1, 0.7 and 2.0 M$_{\oplus}$. We find that except for the case where $M_{\text{pl}} = 0.1\ M_{\oplus}$ (e.g. the case of Mars), all other simulated systems maintain similar flow structures - a stable atmosphere with an ``egg-shape'' region. In the given context, a Mars-sized planet or planetary embryo is not able to hold a regular proto-atmosphere with the little gravity it provides. As can be seen from the leftmost panel in Fig.~\ref{fig:param_pm2D}, the supersonic gas forms a ``secondary shock front'' on the planet surface. 
The planet's gravity is not high enough to focus the incoming gas, i.e. no bow shock is formed, and hence, the incoming gas is able to directly strike onto the planet's surface where it is shocked.
The yellow dashed lines in the figure mark where the gas velocity reduces below sound speed. A planetary embryo with such a small size cannot hold a bound atmosphere.
Recycling vortices are still present behind the planet, though this region can no longer enclose the whole planet. Note that the situation can change with different background set-ups, e.g. reducing the relative velocity between the planet and the disk gas could help a small planet object like Mars to retain a normal atmosphere.
If Mars was once scattered on an eccentric orbit \citep{Hansen2009}, it is likely to experience dramatic atmosphere stripping and re-accretion repeatedly throughout the orbit. 
This can result in the loss of any outgassed atmosphere, which is replaced by the reducing H$_2$/He atmosphere.

On the other hand, more massive planets are able to greatly slow down the supersonic gas and hold atmospheres of a larger mass. The ``egg-shape'' regions around these planets also reveal more complicated vortices and other flow structures.

Fig.~\ref{fig:Param_pm} summaries the trends of atmosphere mass and averaged planet surface pressure with the increasing planet mass, similar to Fig.~\ref{fig:Param_Mach}.
Both properties increase as the planet gets more massive, most dramatically in the sub-Earth mass range. 
In our set-ups, planets with sufficient orbital eccentricity do not possess a bound atmosphere when their mass is lower than 0.5 M$_{\oplus}$. The hydrodynamics around planets with mass $<$ 0.7 M$_{\oplus}$ are dominated by the recycling gas flows.

\subsubsection{The Effect of Background Gas Density} \label{subsubsec:rho_bkg}
The background gas density can have different values from 10$^{-9}$ g cm$^{-3}$ for a number of reasons.
As was mentioned in \S~\ref{subsubsec:bound_init}, different models for the solar/stellar nebula would provide augmented nebular density profiles (0.5 - 1.5 orders of magnitude higher).
Besides the uncertainties from theoretical models, the age of the protoplanetary disk can also affect the nebular density in the background.
Within 1 - 10 Myrs, the nebular gas density can be lowered one to two orders of magnitude as the disk evolves through accretion, wind, and photoevaporation (e.g. \citealt{Mordasini2012, Bai2016, Nakatani2018a, Nakatani2018b}).

\begin{figure*}[ht!]
\centering
\includegraphics[scale=0.255]{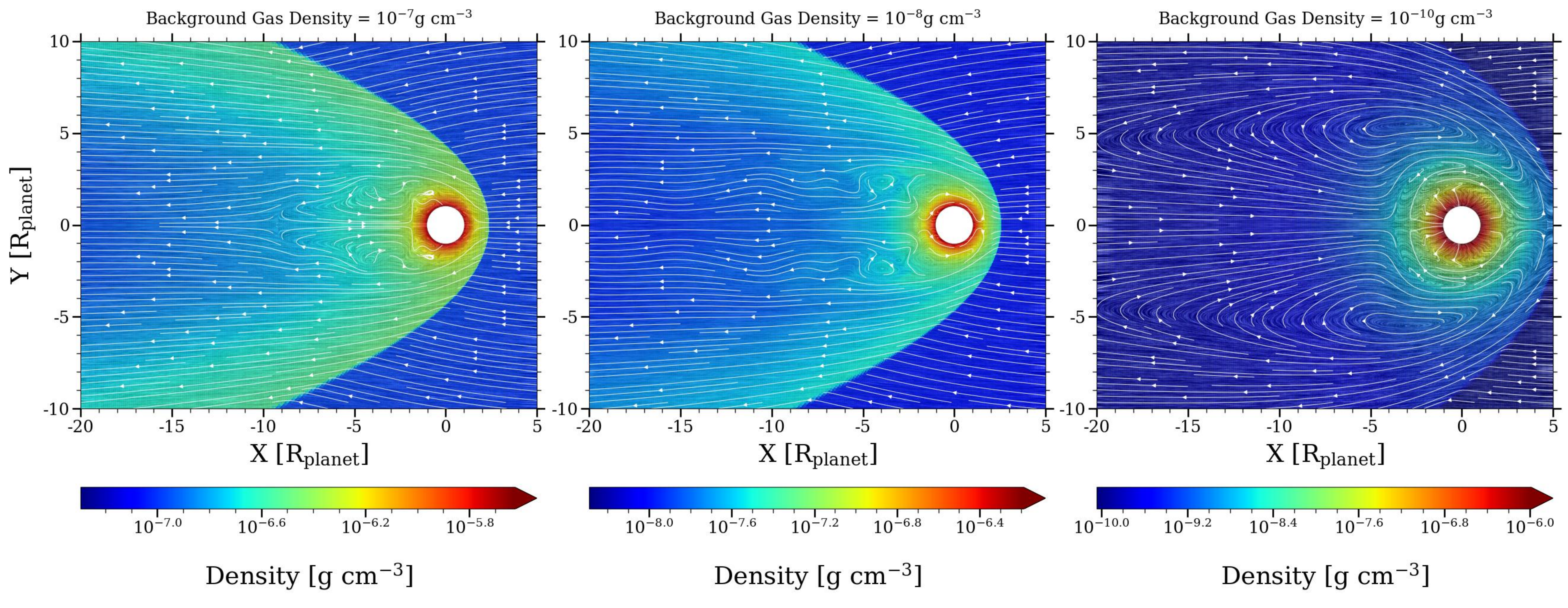}
\caption{Gas density and the flow structure around the planet with various background nebular gas density (10$^{-7}$, 10$^{-8}$ and 10$^{-10}$ g cm$^{-3}$ respectively), illustrated with LIC and streamlines. Note that the color scale in each panel is different from each other. The solutions from the cases of 10$^{-7}$ and 10$^{-8}$ g cm$^{-3}$ show a different flow structure than the regular ``egg-shape'' pattern. Vortices behind the planet are unstable - they constantly develop, get shredded and disintegrate. In particular, Kelvin--Helmholtz instability is observed in the case of 10$^{-8}$ g cm$^{-3}$, forming density waves parallel to the symmetry axis. No gas is being steadily recycled but some atmosphere stays bound to the planet in both cases with the elevated background density. 
\label{fig:param_den2D}}
\end{figure*}

Unlike the relative velocity and the planet mass, the impact from the nebular gas density to our simulation solutions seems less straightforward.
In our study, we perform the simulations with the background gas density varying from 10$^{-7}$~g cm$^{-3}$ to 10$^{-10}$~g cm$^{-3}$ while keeping other parameters the same.
The resulting flow patterns and atmospheric properties do not change monotonically with the nebular density.
When we lower the background gas density, the hydrodynamical solution presents the same flow pattern -- the ``egg-shape'' recycling region (the right panel of Fig.~\ref{fig:param_den2D}). However, the region expands and the planet is able to hold a more massive atmosphere.
The decrease of the nebular density leads to a lower gas pressure in the background. With less ram pressure pushing against the moving planet, the bow shock becomes weaker and facilitates gas accretion onto the planet.
In practice, we decrease the background gas density to as low as 10$^{-11}$ g cm$^{-3}$, in which case the planet retains an even larger atmosphere. However, we exclude the results from this simulation as it converges too slow and becomes too expensive computationally to reach a quasi-steady state.

When we elevate the background density, the hydrodynamics of the system seems to switch to a different regime. The solutions present a different flow structure (the left and middle panels in Fig.~\ref{fig:param_den2D}). We no longer observe the stable recycling of gas through the ``egg-shape'' region. Instead, small-scale vortices constantly emerge and disintegrate behind the planet. In the case of 10$^{-8}$ g cm$^{-3}$ (\texttt{DEN-8}), we observe the presence of Kelvin--Helmholtz instability along the direction parallel to the shearing flow. The instability invokes propagating density waves, which are most prominent at two to eight planet radii behind the planet and weakened further downstream.
In the case of 10$^{-7}$ g cm$^{-3}$ (\texttt{DEN-7}), some portion of gas appears to be flowing back to the planet near the symmetry axis. The formation of large vortices to initiate the recycling is however unstable - different scales of vortices are frequently shredded and remerged. In both cases, no nebular gas is being stably recycled but a portion of atmosphere stays bound to the planet. Nonetheless, we can still define an atmosphere using a suitable density contour for each case\footnote{For \texttt{DEL-7} we use the contour line of 10$^{-6.4}$ g cm$^{-3}$; for \texttt{DEL-8} we use 10$^{-7.4}$ g cm$^{-3}$; for \texttt{DEN-10} we use 10$^{-9.4}$ g cm$^{-3}$.}. As the simulation converges, the atmosphere mass fluctuates around some constant value. By averaging these fluctuating values we have an estimate of the amount of atmosphere around the planet.

The fact that the hydrodynamics change with the background density is an effect of radiative transfer. Because the equations of hydrodynamics do not set any absolute scale of density, the solutions of simulations without radiative transfer should be self-similar, with the background density being a scaling factor.
A typical optical depth for the proto-atmosphere within the bow shock is $\sim$ 200 when the background density is 10$^{-9}$ g cm$^{-3}$.
The quantity reaches $\sim$ 600 when the background density is 10$^{-10}$ g cm$^{-3}$ and drops to $\sim$ 20 -- 30 when the background density is 10$^{-8}$ g cm$^{-3}$ or 10$^{-7}$ g cm$^{-3}$.
The drastic drop in opacity and optical depths of atmospheres in the high-background density cases are caused by the evaporation of dust -- the gas temperatures near the planet surface exceed 1500 K. This leads to a different hydrodynamics regime compared to the recycling regime in the lower-background density cases.

\begin{figure}[ht!]
\centering
\includegraphics[scale=0.28]{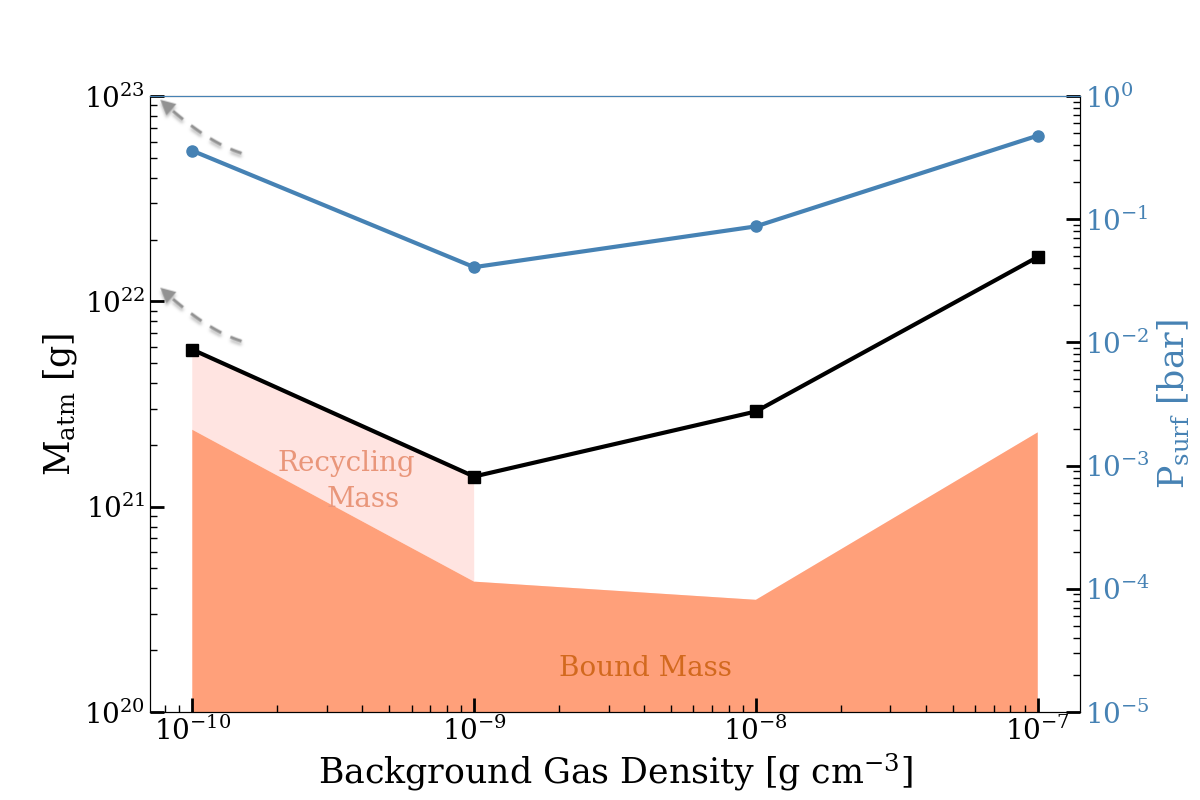}
\caption{Similar to Fig.~\ref{fig:Param_Mach} and Fig.~\ref{fig:Param_pm}. The impact from the background gas density ($\rho_{\infty}$) on the atmosphere mass and averaged planet surface pressure. Both increase with a smaller nebular density due to the lack of ram pressure. The gray dashed arrows indicate the trends as the density gets further reduced. The quantities also increase with a larger nebular gas density as the hydrodynamics switch to a different regime. See the text for the details. \label{fig:Param_den}}
\end{figure}

Fig.~\ref{fig:Param_den} summarizes how the atmosphere mass and surface density change with the background gas density. We observe two hydrodynamics regimes operating in different background-density cases, which are likely invoked by the opacity jump in the atmospheres. 
In either regime, there are two competing mechanisms that affect how much atmospheres can the planet holds.
In the regime where instabilities are present and the large-scale gas recycling is absent (\texttt{DEN-7} and \texttt{DEN-8}), the atmosphere is more massive because of the sufficient gas supply from a dense nebular environment. This factor wins over the stripping effects from a larger ram pressure against the bow shock. The atmosphere is thin and compact.
In the recycling regime, however, the drop in background gas density does not significantly limit the availability of nebular gas for accretion, but rather facilitates gas accretion by weakening the ram pressure. The atmosphere is thick and puffy. We conclude that the recycling regime is favorable for planets to keep a stable atmosphere in the planetary bow shock.

Lastly, we would like to point out that the parameter study in background gas density also provides qualitative estimates to the cases where planets are located at different distances from the star. 
If a planet is on an orbit further away from the star (e.g. 10 au), the decrease in the surrounding gas temperature and density can result in the reduce in the ram pressure against the planetary bow shock. The flow structure of the proto-atmosphere likely resembles that of the low-density case study, where the planet retains a puffy and massive atmosphere in the recycling regime.
On the other hand, a planet located closer to the star is exposed to a denser and hotter nebular environment (e.g. 0.1 au). With the surrounding temperature approaching 1000 K and the nebular density raised by 2 -- 3 orders of magnitude, the proto-atmosphere likely presents instabilities as in those high-density case studies.

\subsection{Orbit-dependent simulation results} \label{subsec:orbit}

In the second part of our hydrodynamic simulations, we focus on tracking the evolution of the proto-atmosphere as the planet moves on an eccentric orbit.
All parameters relevant to the planet--disk environment are time-dependent, including the background gas density, the temperature and the relative velocity between the planet and the gas (only the speed is time-dependent, the velocity direction is assumed to be fixed given that the change of bow shock orientation is negligible in local-framed simulations - Fig.~\ref{fig:Schem_orbit}).
Therefore, the outcomes of the simulation are solely controlled by the orbital parameters (the semimajor axis and the eccentricity) as well as the planet mass. 

\begin{figure*}[ht!]
\centering
\includegraphics[scale=0.65]{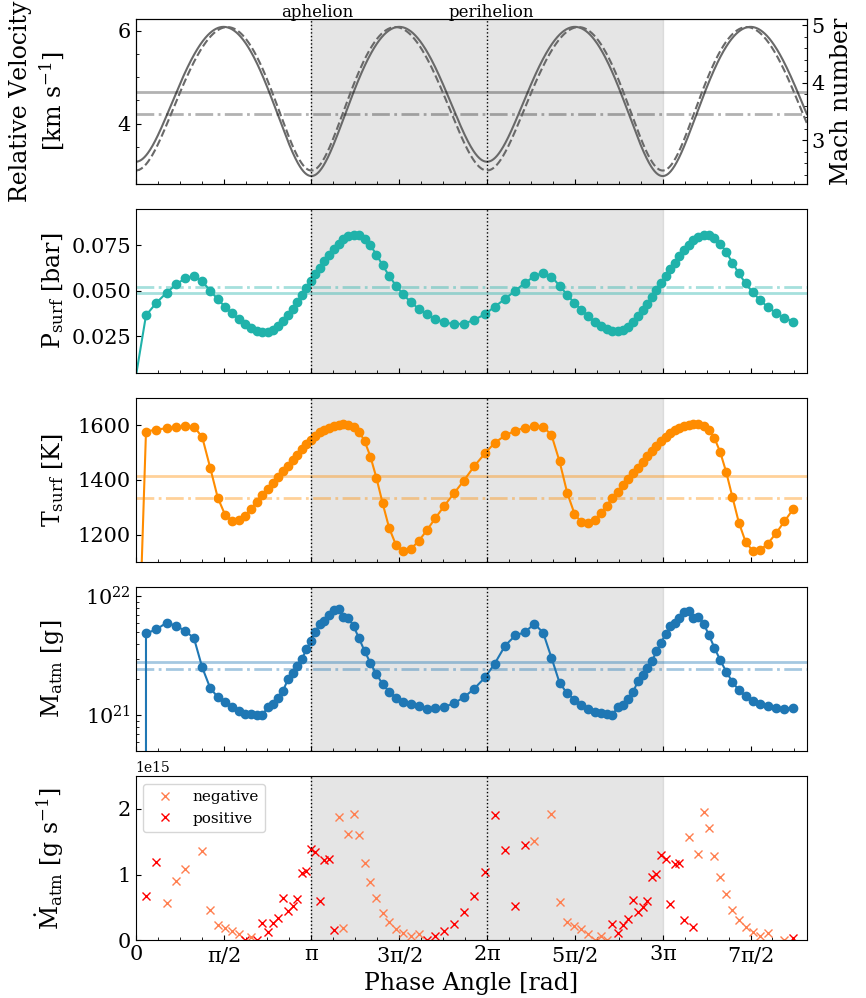}
\caption{The evolution of the proto-atmosphere as a 1 M$_{\oplus}$ planet travels on an eccentric orbit ($a$ = 1 au, $e$ = 0.2). Periodic ``forced oscillations'' are reflected on all atmospheric properties. 
Panel 1: the relative velocity between the planet and the gas (solid line) and the corresponding Mach number (dashed line), same as Fig.~\ref{fig:relative_vel}; Panel 2: the atmospheric pressure on the planet surface; Panel 3: the planet surface temperature; Panel 4: the atmosphere mass defined by density contours; Panel 5: the gas accretion or mass loss rate, the accretion rate is positive (cross markers in red color) and the loss rate is negative (cross markers in coral color). 
The gray shaded region represents a complete orbital period from one aphelion to the next. The vertical dotted lines denote where the planet reaches its aphelion and perihelion.
The horizontal solid line in each panel shows the orbit-averaged value of the corresponding quantity, while the horizontal dot-dashed lines represent the results from the ``snapshot'' simulation \texttt{VEL4.2}.
The proto-atmosphere shows a delay in response to the boundary changes caused by orbital variations. 
The evolution of the proto-atmosphere from aphelion to perihelion and from perihelion to aphelion are not symmetric due to the orbital effects. See text for a detailed discussion.
\label{fig:orb_osc}}
\end{figure*}

We model a planet with 1 M$_{\oplus}$ traveling on an orbit with the semimajor axis $a$ = 1 au and eccentricity $e$ = 0.2. The simulation starts with the planet at its perihelion. 
Fig.~\ref{fig:orb_osc} presents the periodic features reflected on the atmospheric properties as the planet completes 1.9 orbits on the orbit.
At small phase angles ($< \pi/3$), the planet undergoes a fast-accretion stage, trying to reach an equilibrium with the surrounding nebula. 
To ensure that the fluctuations of all calculated quantities are caused merely by orbital effects, we exclude the first half of an orbital period.
From the first aphelion to the second aphelion (phase angle = $\pi$ -- $3\pi$, shaded by light gray in Fig.~\ref{fig:orb_osc}), the planet completes a full orbit.

We can see that the relative velocity between the planet and the gas rises from a minimum as the planet leaves its aphelion. However, the atmospheric properties -- the surface pressure, temperature, the mass, and the mass accretion/loss rate -- do not respond to the change immediately.
The proto-atmosphere should have been undergoing stripping with the increasing relative velocity/Mach number, but the surface pressure, temperature, and mass continue to grow for a while before decreasing. 
On the other hand, the atmosphere is still experiencing mass loss even when the relative velocity starts to decline after reaching a peak (phase angle = $3\pi/4$). 
To sum up, the system shows a delay in response to the time-dependent boundary conditions. The oscillations in atmospheric properties show a phase shift of $\sim \pi/4$, corresponding to an average reaction time $\sim 7\times 10^{6}$~s\footnote{We obtain the lag time by comparing the locations of peaks and valleys of the curves of the surface pressure and the relative velocity (Fig.~\ref{fig:orb_osc}).} that matches the dynamical timescale of the system.

\begin{figure*}[ht!]
\centering
\includegraphics[scale=0.24]{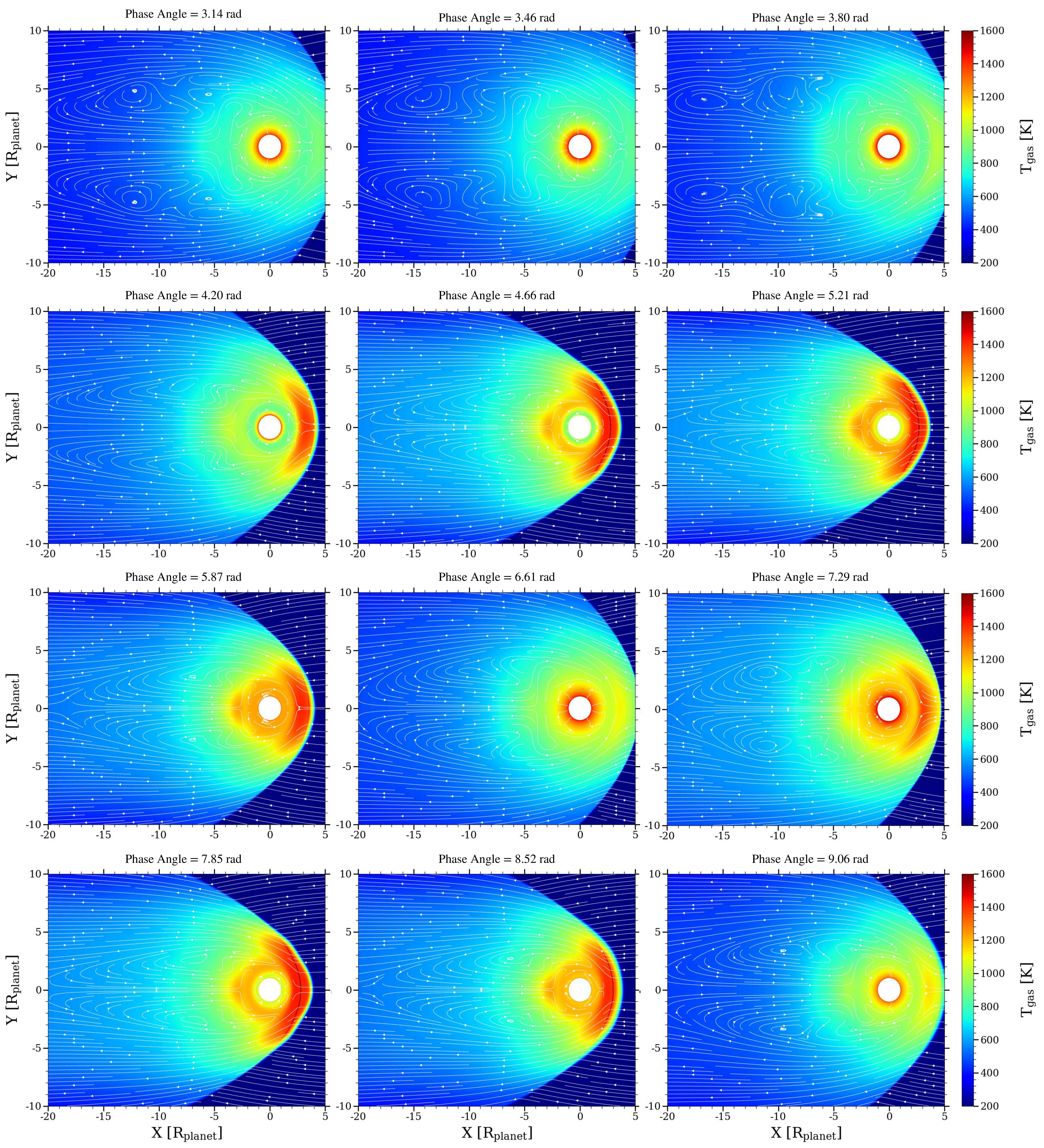}
\caption{The evolution of gas temperature over one full orbit (phase angle $\pi \sim 3\pi$, matching the gray shaded region in Fig.~\ref{fig:orb_osc}). As the relative velocity between the planet and the gas increases, the bow shock reheats the proto-atmosphere from outside to inside, stopping the radiative cooling.
\label{fig:Tgas_orb}}
\end{figure*}

Note that in the orbit-dependent simulation, the relative velocity is not the only parameter affecting how much of the atmosphere can be retained. The time-dependent background gas temperature, density, and pressure can also make a difference, although their variations within one orbit are much smaller compared to the relative velocity. 
For example, the relative velocity/Mach numbers are similarly small for the planet at aphelion and perihelion. But the planet seems to retain more atmosphere after it passes aphelion, because the denser and hotter disk environment at perihelion leads to a larger ram pressure towards the planet, making the gas accretion less effective.
We also notice that the phase changes in the surface temperature are not entirely synchronized with the other properties -- it drops much faster than it rises. Spontaneously, one could expect the surface temperature to change in phase with the change of surface pressure for a pressure-supported atmosphere. 
But when the surface pressure is high, the surface temperature can reach 1500 -- 1600 K and drastically lower the local opacity (dust is evaporated). So as the atmosphere gets stripped (at high relative velocity/Mach number) it undergoes the fast radiative cooling. 
However, the high relative velocity also produces a strong bow shock that dramatically heats up the compressed gas at the shock front. The heat can be transferred to the atmosphere from outside to inside, creating a temporary temperature inversion. Eventually, the heat from shock gas prevents the surface temperature from further decrease and causes it to rise again (Fig.~\ref{fig:Tgas_orb}).

The orbit-averaged values of different quantities -- drawn as horizontal solid lines in Fig.~\ref{fig:orb_osc} -- are consistent with the results acquired from the ``snapshot'' simulation \texttt{VEL4.2} (horizontal dot-dashed lines).
With the appropriate selection of input parameters, the ``snapshot'' simulation is also able to represent the ``averaged'' picture of the dynamic proto-atmosphere over the corresponding orbit.

\section{Discussions}  \label{sec:discussions}

\subsection{Comparison with Planets on Circular Orbits} \label{subsec:compare_circ}
The proto-atmosphere on low-mass planets with circular orbits has been studied in detail to explore the formation of super-Earths and Earth-like planets (see literature review in \S~\ref{sec:intro}).
By comparing our results with the outcomes from hydrodynamic simulations for planets on circular orbits, we identify a few major differences in atmosphere accretion on planets with the two types of orbits.

1) The flow structure. 
The relative motion between the planet and gas is subsonic in cases where planets are on circular orbits. The gas accretion does not generate a bow shock.
In 1D hydrodynamic simulations (e.g. \citealt{Stokl2015, Stokl2016}), proto-atmosphere accretion on the planet is assumed to be in azimuthal symmetry -- the infall of the nebular gas lies in the radial direction.
In 2D and 3D simulations (e.g. \citealt{Ormel2015a, Ormel2015b, Cimerman2017, Kuwahara2019, Fung2015}), the planet is considered to be embedded in a Keplerian disk. A local shearing flow with its reference frame is comoving with the planet as the gas travels faster on the side facing the star and slower on the other side. Both 2D and 3D simulations find a pressure-supported rotating atmosphere around the planet. In 2D, the atmosphere becomes isolated from the disk gas when the system reaches a steady state. In 3D, the atmosphere stays connected to the disk as the atmosphere is constantly exchanging gas with the nebula. The incoming gas enters the Bondi sphere around the planet at high latitudes and leaves at the equatorial regions. In all simulations, two horseshoe flow regions are seen on the front and rear sides of the planet.
The situation of a planet experiencing a headwind from the disk gas in sub-Keplerian motion has also been considered. In such cases, the flow pattern becomes asymmetric around the planet as the streamlines are compressed towards the headwind. The nebular gas enters from the equatorial region and leaves at high latitudes.
For the detailed flow pattern figures of the above simulations, please refer to \cite{Ormel2015a, Ormel2015b, Cimerman2017, Kuwahara2019, Fung2015}.
In summary, proto-atmospheres on planets with both circular (subsonic) and eccentric (supersonic) orbits are actively interacting with the nebular gas through large-scale recycling. The recycling/replenishment timescale in the circular orbit case is generally longer ($> 10^7$~s), compared to that of a typical eccentric planet simulated in this study (5$\times$10$^5$ - 5$\times$10$^6$~s).

2) The steady state. In 2D simulations for planets on circular orbits, the gas accretion seems to be able to reach a steady state where the solutions no longer fluctuate. The flow pattern and other physical quantities (e.g. density, velocity) could be, however, time-variable in some 3D simulations \citep{Cimerman2017}. But some other simulations suggest that they can reach steady states (Moldenhauer et al., in prep.).
In comparison, gas accretion onto planets with eccentric orbits can also reach steady/quasi-steady states as shown in this study. 
   
3) Atmospheric properties. Planets on circular orbits are able to hold much more massive atmospheres. The 1D models in \cite{Stokl2015, Stokl2016} points out that a 1 M$_{\oplus}$ planet can accrete up to 300 bars (or 10$^{25}$~g) of atmosphere, which is $\sim$ 0.2\% of the planet mass. The 2D simulations in \cite{Ormel2015a} and 3D models (Moldenhauer et al., in prep.) give similar or slightly lower values. 
In terms of surface temperature, a 1 M$_{\oplus}$ planet on a circular orbit can be $\geq$ 3000 K (Moldenhauer et al. in prep. \& \citealt{Stokl2016}), whereas for an eccentric orbit we find 1000 -- 2000 K. This is due to the accretion of a much smaller atmosphere and implies dust grains dominate the opacity in such an atmosphere.

A summary of the above comparison is presented in Table \ref{tab:comp_circ}.

\begin{table*}[]
    \centering
    \begin{tabular}{M{4.5cm}M{4.5cm}M{4.5cm}}
    \hline
    \hline
    Orbit Type  &  Circular  & Eccentric (2D)   \\  \hline
    Flow Structure  &  2D: Atmosphere is isolated;\newline 3D: Atmosphere exchanges gas with the nebula   &  Atmosphere exchanges gas with the nebula   \\
    Steady State   &   2D: yes; 3D: no   &   yes   \\
    Surface Pressure (1 M$_{\oplus}$)   &   10$^2$ -- 10$^{2.5}$ bar  &  10$^{-2}$ -- 10$^{-1}$ bar  \\
    Surface Temperature (1 M$_{\oplus}$) &  $\geq$ 3000 K    &   1000 -- 2000 K  \\
    Atmosphere Mass (1 M$_{\oplus}$) &   10$^{24}$ -- 10$^{25}$ g\newline (10$^{-3}$ -- 10$^{-2}$ M$_{\oplus}$)   &  10$^{21}$ -- 10$^{22}$ g\newline (10$^{-6}$ -- 10$^{-5}$ M$_{\oplus}$) \\  \hline
    \end{tabular}
    \caption{Brief comparison of hydrodynamic simulations for proto-atmospheres on planets with circular orbits and eccentric orbits}
    \label{tab:comp_circ}
\end{table*}{}

\subsection{Comparison with the Bondi-Hoyle-Lyttleton Accretion} \label{BHL}

The accretion on a gravitating object traveling supersonically in a homogeneous medium is a classical problem called the Bondi-Hoyle-Lyttleton (BHL) accretion \citep{Hoyle1939, Bondi1944}.
For decades, numerous analytical and numerical studies have been exploring the flow structure and the stability of BHL accretion (see \cite{Edgar2004, Foglizzo2005} for reviews). 
Typical applications for these BHL accretion models include wind accretion in binary systems (e.g. \citealt{Taam1991, Xu2019}), accretion on other compact objects like black holes (e.g. \citealt{Pogorelov2000}), and the proto-stellar clusters and galaxy clusters (e.g. \citealt{Bonnell2001, Sakelliou2000}). 

Gas accretion on an eccentric planet is also a BHL accretion problem in principle, although the complications from radiative properties and boundary conditions can deviate the solutions from the classical derivatives.
In fact, the flow structure and stability of BHL accretion are proven strongly dependent on the chosen grid method (2D planar, 2D axisymmetric or 3D), the given parameters and the accretor's properties \citep{Foglizzo2005}. 
Our results in this study are aligned with the previous finding that simulations in 2D axisymmetric are in general stable (e.g. \citealt{Pogorelov2000, Foglizzo2005}).
The ``egg-shape'' flow pattern -- where a portion of the gas flow back to the planet in the post-shock region -- is also commonly observed in other axisymmetric simulations (e.g. \citealt{Koide1991, Pogorelov2000, MacLeod2015}). 

\subsection{The Damping of Orbital Eccentricity}

Though planets and planetary embryos can be scattered and/or migrate in the young protoplanetary disk, their phases on eccentric orbits are believed to be temporary as long as the disk gas is present.
Planets traveling in a gaseous disk can experience both aerodynamical and gravitational gas drag.
The aerodynamical drag is the pressure resistance that any gas or fluid exerts on the object moving inside it, and it is more effective for smaller bodies like planetesimals (e.g. \citealt{Kominami2002}).
The gravitational drag arises from torques as the planet interacts with the gaseous disk (e.g. \citealt{Papaloizou2000}).
It is more prominent in massive objects like planetary embryos and planets.


The aerodynamical damping timescale can be calculated from the formulae in \cite{Tanaka1999, Kominami2002} as:
\begin{equation} \label{eqn_t_aero}
\begin{aligned}
        \tau_{\text{aero}} \approx \frac{M_{\text{pl}}}{\pi {R_{\text{pl}}}^2\rho_{\infty}v_{\infty}} \approx 3\times 10^5 \left(\frac{e}{0.2}\right)^{-1} \left(\frac{M_{\text{pl}}}{M_{\oplus}}\right)^{1/3} \\ \times \left(\frac{\rho_{\infty}}{10^{-9} \text{g}\ \text{cm}^{-3}}\right)^{-1} \left(\frac{r}{1\ \text{au}}\right)^{1/2} \text{yr},
\end{aligned}{}
\end{equation}{}
where we approximate the relative velocity between the planet and the gas $v_{\infty}$ as $(1/2)ev_{\text{K}}$, $v_{\text{K}} = (GM_{\odot}/r)^{1/2}$ is the Keplerian speed.


For low eccentricities (e.g. $e \leq H/r$ or $v_{\infty} \leq c_{\infty}$), the damping timescale for the gravitational drag is given as:
\begin{equation} \label{eqn_t_grav1}
    \begin{aligned}
        \tau_{\text{grav,le}} \approx \left(\frac{M_{\odot}}{M_{\text{pl}}}\right) \left(\frac{M_{\odot}}{\Sigma_{\infty}r^2}\right) \left(\frac{c_{\infty}}{v_{\text{K}}}\right)^4 \Omega_{\text{K}}^{-1} \\
        \approx 5\times 10^2 \left(\frac{M_{\text{pl}}}{M_{\oplus}}\right)^{-1} \left(\frac{\rho_{\infty}}{10^{-9} \text{g}\ \text{cm}^{-3}}\right)^{-1} \\ \times \left(\frac{r}{1\ \text{au}}\right)^{-3/4} \text{yr}.
    \end{aligned}{}
\end{equation}{}
$\Sigma_{\infty} = \rho_{\infty}H\sqrt{2\pi}$ is the disk gas surface density. $\Omega_{\text{K}} = (GM_{\odot}/r^3)^{1/2}$ is the Keplerian frequency \citep{Kominami2002}.

However, \cite{Papaloizou2000} found that when $e > 1.1H/r$, the sign of the torque on the planet reverses and leads to an eccentricity damping timescale strongly dependent on $e$. The disk scale height-to-radius ratio $H/r$ ranges from 0.02 to 0.07 throughout the protoplanetary disk (0.1 - 10 au). For planets in the supersonic regime as we consider in this study ($e \geq 0.1$), we need to calculate a damping timescale different from Eqn. \ref{eqn_t_grav1}. For such cases, the authors suggest to replace $\tau_{\text{grav,le}}$ to $[e/(H/r)]^3\tau_{\text{grav,le}}$. So we have:
\begin{equation} \label{eqn_t_grav2}
\begin{aligned}
    \tau_{\text{grav,he}} \approx 1\times10^5 \left(\frac{e}{0.2}\right)^3 \left(\frac{M_{\text{pl}}}{M_{\oplus}}\right)^{-1} \left(\frac{\rho_{\infty}}{10^{-9} \text{g}\ \text{cm}^{-3}}\right)^{-1} \\ \times \left(\frac{r}{1\ \text{au}}\right)^{-3/2} \text{yr}.
\end{aligned}
\end{equation}{}

The ratios of the these timescales are:
\begin{equation} \label{eqn_t_ratio1}
    \begin{aligned}
    \frac{\tau_{\text{aero}}}{\tau_{\text{grav,le}}} \approx 600\times \left(\frac{e}{0.2}\right)^{-1} \left(\frac{M_{\text{pl}}}{M_{\oplus}}\right)^{4/3} \left(\frac{r}{1\ \text{au}}\right)^{5/4}
    \end{aligned}
\end{equation}{}

\begin{equation} \label{eqn_t_ratio2}
    \begin{aligned}
    \frac{\tau_{\text{aero}}}{\tau_{\text{grav,he}}} \approx 3\times \left(\frac{e}{0.2}\right)^{-4} \left(\frac{M_{\text{pl}}}{M_{\oplus}}\right)^{4/3} \left(\frac{r}{1\ \text{au}}\right)^2
    \end{aligned}
\end{equation}{}

Therefore, for planets more massive than 10$^{25}$~g on orbits with low eccentricities, they are mainly circularized by the gravitational drag \citep{Kominami2002}. For planets on orbits with higher eccentricities, aerodynamical gas drag dominates if $M_{\text{pl}} <$ 0.1 M$_{\oplus}$, while gravitational drag dominates if $M_{\text{pl}} >$ 1 M$_{\oplus}$. These calculations are consistent with calculations of eccentricity damping by \cite{Morris2012}.

The orbital eccentricity can also be damped by the dynamical friction coming from the swarms of planetesimals scattered throughout the disk. The damping timescale has a similar expression as Eqn. \ref{eqn_t_grav1}, except $\Sigma_{\infty}$ is substituted by the surface density of solids in the disk $\Sigma_{\text{solid}}$ and $c_{\infty}$ is substituted by the velocity dispersion of planetesimals $v_{\text{dis}}$. Therefore, the mechanism can either be incorporated into the gravitational drag regime \citep{Kominami2002}, or neglected in cases where $e > 0.03$ as the gravitational drag dominates over dynamical friction \citep{Morris2012}.

The damping of orbital eccentricity follows
\begin{equation}
    \frac{de}{dt} = - e\tau_{\text{damp}}^{-1}
\end{equation}{}
where $\tau_{\text{damp}}$ is the damping timescale. If substituted by $\tau_{\text{grav,he}}$, $de/dt$ becomes linearly proportional to $e^{-2}$ and M$_\text{pl}$. The lower the eccentricity (still larger than 0.1) and the larger the planet mass, the faster its orbit gets damped. 

For a typical planet simulated in our models, its eccentric orbit is circularized in about 0.1 Myr, which is much longer than the relevant timescales in this study. The damping of eccentricity is therefore negligible in both ``snapshot'' and orbit-dependent simulations.
The planets may be able to hold more atmospheres as their orbits get circularized if the ambient disk gas has not yet dissipated. But this procedure can be complicated, because the damping also vanishes with the disappearing disk gas.

\subsection{Limitations of This Study}
The study does not consider the cases where the relative velocity between the planet and the gas is below sound speed (e.g. orbital eccentricity is very small, $e < 0.08$). In such cases, the planet does not generate a bow shock that compresses and heats up the gas in front of the planet, and the stripping of the proto-atmosphere will be much less effective.
In fact, it is no longer suitable to model planets in this regime with 2D axisymmetric simulations because the relative velocity ($v_{\infty}$) could be comparable to the shear or headwind velocity of the ambient gas.
To model gas accretion on planets traveling at subsonic speed, we suggest using the grid methods adopted for planets experiencing a headwind on circular orbits (\citealt{Ormel2015a, Ormel2015b}, MMoldenhauer et al., in prep.). The proto-atmosphere is predicted to be asymmetric and less massive compared to the circular cases, but much more massive than the supersonic cases. 
As our paper solely focuses on eccentric planets in the supersonic regime, the modeling of gas accretion on subsonic planets is left as future work.
It will be interesting to compare the results from these models with the circular cases and the supersonic cases.

Another limitation of this study comes from the axisymmetry assumption. A number of numerical simulations on BHL accretion have adopted the full 3D grid method to represent the more realistic geometry. Unfortunately, most of them are not stable \citep{Foglizzo2005}. The 3D simulations usually break the axial symmetry of the flow structure and develop instabilities in the downstream of the bow shock (e.g. \citealt{Ruffert1994a, Ruffert1994b}). These instabilities are not as prominent and strong as the ``flip-flop'' behaviors observed in 2D planar simulations and are possible to arise from numerical artifacts.
Nonetheless, the gas is still observed to be flowing back towards the accretor in the post-shock region, similar to the 2D axisymmetric simulations. We expect that the hydrodynamical solutions for the atmospheres (e.g. density, pressure, temperature, mass) will be within the same orders of magnitude as obtained in this study, regardless of the adopted grid methods.

Lastly, we would like to point out that we neglect solid accretion that could occur at the same time as gas accretion. We assume that the planet never grows in mass and the planet luminosity is barely affected by pebble/planetesimal accretion. Solid accretion is common and important for planet formation in young disks ($\le$ 5 Myr), and luminosity feedback could be important \citep{Mordasini2013}. 
There are several studies that have considered the impacts of orbital eccentricity on solid accretion (e.g. \citealt{Lissauer1997, Liu2018}).
Given that the timescales relevant to gas accretion in this study are days to weeks, the neglect of long-term effects from solid accretion can be justified.

\section{Summary}   \label{sec:summary}
Planet migration and scattering are common in young planetary systems. Planets can be excited onto eccentric orbits and their atmospheres can undergo drastic stripping.
In this study, we have performed hydrodynamic simulations to explore the presence of dynamic H$_2$/He proto-atmospheres around low-mass planets with eccentric orbits. We have especially focused on planets that are traveling supersonically through the disk gas.
In the ``snapshot'' simulations, we specify the disk--planet environment with the relative velocity between the planet and the gas, the planet mass and the background gas density as input parameters. We analyze the corresponding flow structures and discuss how the atmospheric properties depend on the change of each parameter.
In the orbit-dependent simulation, we allow the above parameters to be dependent on the time-varying location of the planet on an eccentric orbit to track the orbital evolution of the proto-atmosphere.

Our major findings and conclusions are summarized as follows:
\begin{enumerate}
    \item The hydrodynamic simulations of gas accretion on supersonic planets are generally able to reach steady states. Gas entering the planetary bow shock with small impact parameters flow back towards the planet in the downstream of the shock, forming an ``egg-shape'' region with stable vortices. The proto-atmosphere is exchanging gas with the nebula via the large-scale recycling within this region, while a small portion of the gas can stay bound to the planet. The recycling timescale is estimated to be 10$^5$ - 10$^7$~s depending on the actual disk-planet environment.
    
    \item Planets on eccentric orbits are able to retain atmospheres despite the supersonic speeds relative to the nebular gas. But they possess much less proto-atmospheres compared to planets on circular orbits. For the investigated planets traveling at supersonic speeds, the stripping effect from the planetary bow shock is so strong that the atmospheric surface pressures are merely 10$^{-2}$ to 10$^{-1}$ bar, about four orders of magnitude lower than those of planets on circular orbits. The range of the atmosphere mass is 10$^{21}$ -- 10$^{22}$~g (10$^{-5}$ -- 10$^{-4}$\% M$_{\oplus}$), which is about three to four orders of magnitude smaller than that of the circular cases. Therefore, planet migration and/or scattering can be recognized as one of the important causes of atmosphere stripping and an efficient way of mass loss.
    
    \item The larger the relative velocity between the planet and the gas is, the smaller the proto-atmosphere becomes. High relative velocity/Mach number drives stronger planetary bow shocks that effectively strip the atmospheres. 
    \item The more massive the planet is, the larger the amount of atmosphere it can hold. Planets with a mass lower than 0.5 M$_{\oplus}$ cannot hold a bound atmosphere, but the situation can change with different background set-ups. 
    If a planet with such low mass is ever on an eccentric orbit, it likely also loses any outgassed atmosphere and replaces that with a nebular atmosphere. Such process has chemical consequences by exposing the planet to much more reducing gases (H$_2$ vs. CO$_2$, H$_2$O).
    
    \item The planet could retain more atmosphere when the background gas density is lower (e.g. $< 10^{-9}$ g cm$^{-3}$), because of the lower ram pressure coming from the nebula. Augmented gas density ($> 10^{-9}$ g cm$^{-3}$) could, however, introduce instabilities in the flow structure and switch the hydrodynamics to a different regime. In this regime where the recycling flow is no longer present, the planet can also keep more atmosphere because the dense nebula supplies sufficient gas for accretion. 
    
    \item The atmospheric properties (surface pressure, surface temperature and mass) show oscillatory patterns as the planet moves on an eccentric orbit. The time-varying relative velocity between the planet and the gas is the main driver of the oscillations, but the changes in background gas density, temperature and pressure also affect them. The oscillations show a delay in response to the time-dependent boundary conditions, with an average reaction time similar to the dynamical timescale of the system.
    The orbital ``averaged'' picture of the dynamic proto-atmosphere can be represented by the ``snapshot'' simulation.
\end{enumerate}{}

\acknowledgements 

The authors would like to thank the anonymous reviewer for the helpful comments to improve the paper.
This work has benefited from discussions with many colleagues. We thank Tobias Moldenhauer for sharing the outputs from atmosphere simulations of planets with circular orbits for comparison with our results. We thank Aaron C. Boley, Mark Richardson and Zhaohuan Zhu for their generous help in hydrodynamics code selection and their insights in hydrodynamical modeling.
The authors acknowledge Research Computing at Arizona State University for providing High-Performance Computing (HPC) resources that have contributed to the research results reported within this paper.
This work was supported by NASA Headquarters under the NASA Earth and Space Science Fellowship, grant NNX16AP50H.
The results reported herein benefited from collaborations and/or information exchange within NASA’s Nexus for Exoplanet System Science (NExSS) research coordination network sponsored by NASA’s Science Mission Directorate, grant NNX15AD53G (PI S. J. Desch).
CM would like to thank the support from grant DU 414/21-1 SPP 1992 (PI C. Dullemond) that makes the collaboration with RK, G-DM and CD possible.
RK acknowledges financial support via the Emmy Noether Research Group on Accretion Flows and Feedback in Realistic Models of Massive Star Formation funded by the German Research Foundation (DFG) under grant no. KU 2849/3-1 and KU 2849/3-2.
G-DM acknowledges the support of the DFG priority program SPP~1992 ``Exploring the Diversity of Extrasolar Planets'' (KU 2849/7-1) and the support from the Swiss National Science Foundation under grant BSSGI0$\_$155816 ``PlanetsInTime''.

\appendix
\section{Estimation of the Final Atmosphere Mass}\label{append:steady}

\begin{figure*}[ht!]
\centering
\includegraphics[scale=0.25]{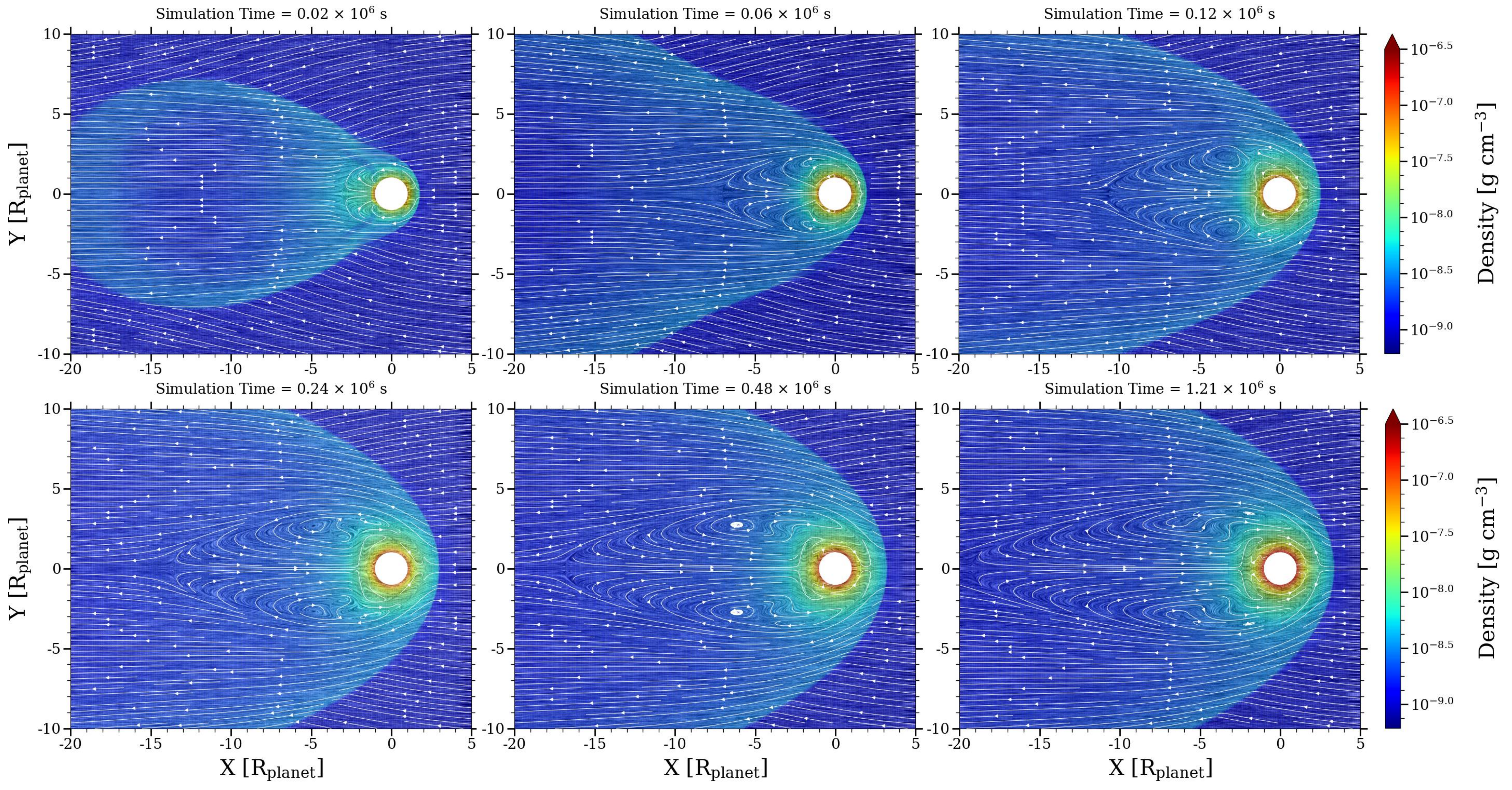}
\caption{The evolution of gas density and the flow structure at the early stage of the canonical simulation. The ``egg-shape'' pattern is formed and stabilized soon after the bow shock is established. Both the proto-atmosphere and the ``egg-shape'' region quickly expand in size over time (see the definitions of proto-atmospheres in \S~\ref{subsubsec:atm_mass}). \label{fig:flow_time}}
\end{figure*}

In our performed simulations, the temporal growth of the proto-atmosphere is rapid in the beginning but the growth rate decreases over time until a steady state is reached.
The flow patterns become stable soon after the bow shocks establish, with the ``egg-shape'' regions and the proto-atmospheres expanding over time. The six snapshots in the canonical simulation in Fig.~\ref{fig:flow_time} show how the flow patterns evolve at the early stage.
Generally, the final atmosphere mass is obtained after the simulated system reaches a steady state. However, in the cases where reaching the steady state takes a great amount of time, it is reasonable to use a fitting model to estimate the final atmosphere mass, and other relevant quantities (e.g. final atmospheric pressure/density on the plant surface).

\begin{figure}[ht!]
\centering
\gridline{\fig{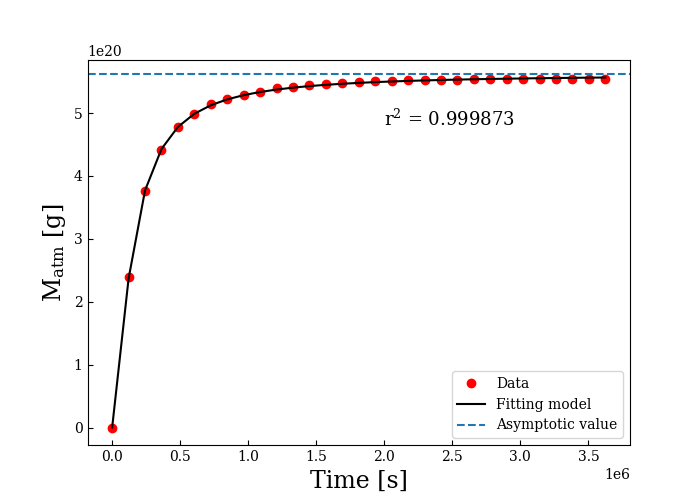}{0.5\textwidth}{(a) Simulation: \texttt{VEL7.0}}
         \fig{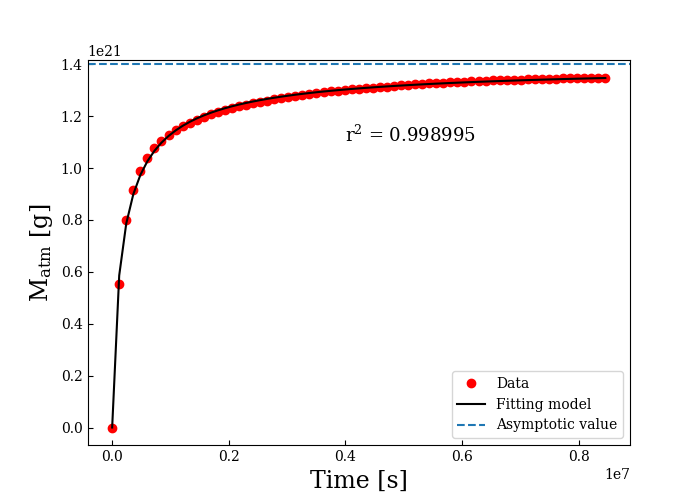}{0.5\textwidth}{(b) Simulation: \texttt{Canonical}}
          }
\caption{The mass of the proto-atmosphere as a function of time (red markers), the fitted growth curve (black solid lines) and the estimated final mass from the fitted asymptotic values (blue dashed lines) in two example simulation cases - a: \texttt{VEL7.0} (relative velocity 7.0 km s$^{-1}$); b: \texttt{Canonical} (relative velocity 4.8 km s$^{-1}$). The coefficient of determination (r$^2$) is denoted for each fitting case.\label{fig:M_time}}
\end{figure}

We adopted an asymptotic model to fit the growth of the atmosphere mass and use the asymptotic value as an estimation of the approximate final mass (Fig.~\ref{fig:M_time}):
\begin{equation} \label{Eq_fit}
    M(t)=a \cdot \frac{t^n}{t^n+b}
\end{equation}
where $t$ is time, $a, b$ and $n$ are the fitting parameters. As $t$ increases, the model infinitely approaches the value of parameter $a$ -- the asymptotic value. 

We perform the line fitting with the function \texttt{\detokenize{curve_fit}} under the \texttt{scipy.optimize} module in \texttt{Python2.7}. The \texttt{\detokenize{curve_fit}} function takes in the fitting model (Eqn. \ref{Eq_fit}), the simulation time $t$, the calculated atmosphere mass from actual outputs and initial guess for the fitting parameters. For each simulation case that needs line fitting, the initial guess for $b$ and $n$ are 20 and 1. We adopt the atmosphere mass calculated from the last output as the initial guess for parameter $a$. The \texttt{\detokenize{curve_fit}} function returns the optimal values of the fitting parameters and the estimated convariance of these values using the least squares regression method. 
The fitting of the above asymptotic model is first tested in simulations that quickly converge and is proven valid in providing the estimated atmosphere mass matching the ones given by the solutions in the steady state (Fig.~\ref{fig:M_time}a). For simulation cases that are run before they reach their steady states (i.e. in quasi-steady states), the obtained asymptotic values from model fitting are used as the approximate estimations of the final atmosphere mass (Fig.~\ref{fig:M_time}b)

Note that here we define the atmosphere as the gas enclosed by the 10$^{-8.5} $~g cm$^{-3}$ density contour. We use this as the default definition unless the background gas density is changed (e.g. in the parameter study). See \S~\ref{subsubsec:atm_mass} for the definitions of proto-atmosphere in detail.




\end{document}